\pdfoutput=1
\documentclass[aps,pre,twocolumn,twoside,showpacs,superscriptaddress]{revtex4}
\usepackage{graphicx,amssymb,amsmath,epsf,bm}
\usepackage{color}

\newcommand{\e}{\mathrm{e}}
\newcommand{\ri}{\mathrm{i}}
\newcommand{\rd}{\mathrm{d}}

\renewcommand{\[}{\left[}
\renewcommand{\]}{\right]}
\newcommand {\be} {\begin{equation}}
\newcommand {\ee} {\end{equation}}
\newcommand {\Be}{\begin{eqnarray*}}
\newcommand {\Ee} {\end{eqnarray*}}
\newcommand {\bey} {\begin{eqnarray}}
\newcommand {\eey} {\end{eqnarray}}
\newcommand{\bit}{\begin{itemize}}      
\newcommand{\eit}{\end{itemize}}
\newcommand{\bfl}{\begin{flusleft}}
\newcommand{\efl}{\end{flusleft}}
\newcommand{\bfr}{\begin{flushright}}
\newcommand{\ec}{\end{center}}
\newcommand{\ben}{\begin{enumerate}}    
\newcommand{\een}{\end{enumerate}}

\newcommand{\comment}[1]{}

\newcommand{\pref}[1]{(\ref{#1})}
\newcommand{\figref}[1]{Fig.~\ref{#1}}

\begin{document}

\title{Chaos in the Hamiltonian mean field model}
\author{Francesco Ginelli}
\affiliation{Istituto dei Sistemi Complessi, CNR, via dei Taurini 19, I-00185 Roma, Italy}
\affiliation{Institute for Complex Systems and Mathematical Biology, King's College, University of Aberdeen, Aberdeen AB24 3UE, United Kingdom}
\author{Kazumasa A. Takeuchi}
\affiliation{Service de Physique de l'\'Etat Condens\'e, CEA . Saclay, F-91191 Gif-sur-yvette, France}
\affiliation{Department of Physics,\! The University of Tokyo,\! 7-3-1 Hongo,\! Bunkyo-ku,\! Tokyo 113-0033,\! Japan}
\author{Hugues Chat\'e}
\affiliation{Service de Physique de l'\'Etat Condens\'e, CEA . Saclay, F-91191 Gif-sur-yvette, France}
\author{Antonio Politi}
\affiliation{Istituto dei Sistemi Complessi, CNR, via Madonna del Piano 10, I-50019 Sesto Fiorentino, Italy}
\affiliation{Institute for Complex Systems and Mathematical Biology, King's College, University of Aberdeen, Aberdeen AB24 3UE, United Kingdom}
\affiliation{Centro Interdipartimentale per lo Studio delle Dinamiche Complesse, via Sansone, 1 - I-50019 Sesto Fiorentino, Italy}
\author{Alessandro Torcini}
\affiliation{Istituto dei Sistemi Complessi, CNR, via Madonna del Piano 10, I-50019 Sesto Fiorentino, Italy}
\affiliation{Centro Interdipartimentale per lo Studio delle Dinamiche Complesse, via Sansone, 1 - I-50019 Sesto Fiorentino, Italy}
\affiliation{INFN Sez. Firenze, via Sansone, 1 - I-50019 Sesto Fiorentino, Italy}

\date{\today}

\begin{abstract}
We study the dynamical properties of the canonical ordered phase of the Hamiltonian mean-field (HMF) model, 
in which $N$ particles, globally-coupled via pairwise attractive interactions, form a rotating cluster.
Using a combination of numerical and analytical arguments, we first
show that the largest Lyapunov exponent
remains strictly positive in the infinite-size limit, converging to its asymptotic value with $1/\ln N$ corrections.
We then elucidate the scaling laws ruling the behavior of this asymptotic value in the critical region
separating the ordered, clustered phase and the disordered phase present at high energy densities.
We also show that the full spectrum of Lyapunov exponents consists of a bulk component converging
to the (zero) value taken by a test oscillator forced by the mean field, plus subextensive bands of
${\cal O}(\ln N)$ exponents taking finite values. We finally investigate the robustness of these results
by studying a ``2D'' extension of the HMF model where each particle is endowed with 4 degrees 
of freedom, thus allowing the emergence of chaos at the level of single particle.
Altogether, these results illustrate the subtle effects of global (or long-range) coupling, and the
importance of the order in which the infinite-time and infinite-size limits are taken: for an infinite-size
HMF system represented by the Vlasov equation, no chaos is present, while chaos exists
and subsists for any finite system size.
\end{abstract}
\maketitle
\section{Introduction}
\label{sec_intro}

In the context of Hamiltonian systems with long-range interactions, the
Hamiltonian mean-field (HMF) model introduced 
  independently in the nineties by Antoni and Ruffo \cite{hmf}
and by Kaneko, Konishi and Inagaki \cite{J-hmf} became the main benchmark for the investigation of thermodynamic
and dynamical properties of non-additive systems \cite{rev1,rev2}. The HMF model
describes an ensemble of $N$ particles moving on a circle, coupled by
pairwise (sinusoidal) attractive interactions. Each particle can also be seen as
a pendulum in a fluctuating potential, whose amplitude is determined
self-consistently and corresponds to the magnetization \cite{hmf}. Detailed
studies of the HMF model have revealed unusual properties, such as ensemble
inequivalence (associated with the occurrence of negative specific heat),
long-lived quasi-stationary states, and anomalous diffusion \cite{rev1,rev2}.
Here, we are interested in the dynamical properties of the standard (microcanonical) 
equilibrium phases. Below the critical energy $U_{\rm c}=3/4$, the HMF system has a finite
magnetization (clustered phase), while above $U_c$ the magnetization vanishes
(homogeneous phase). The two regimes are separated by a second order canonical
phase transition. Both in the limit
 $U \to 0$ and $U \to \infty$
 the dynamics is
integrable: in the former case, all particle are trapped in the (harmonic)
bottom part of the potential well; in the latter, they move freely along
the circle. At intermediate energies, the (nonlinear) microcanonical dynamics 
of a finite system made of $N$ particles is characterized by a spectrum of
Lyapunov exponents (LEs) $\{ \lambda_i \}$ with $ i=1, \dots , 2 N$ and, due to
the Hamiltonian structure, $\lambda_i = - \lambda_{2N+1-i}$.

The thermodynamic limit is, however, a rather intriguing subject. For
$N \to \infty$, the mean field is constant and the evolution of each particle
is equivalent to the motion of a standard pendulum in a constant gravitational
field. Accordingly, no chaos but just periodic orbits can be generated. This
straightforward theoretical prediction is consistent with numerical simulations 
in the homogeneous phase, where it is numerically observed \cite{rev1,latora98}
 that the maximal LE $\lambda_1$ vanishes as $N^{-1/3}$
(a result which can be easily
explained \cite{rev1,Anteneodo_Vallejos-PRE2001} by invoking arguments developed for products of random
matrices \cite{random}).
In contrast, in the clustered phase,
some numerical investigations suggest
that $\lambda_1$ remains finite in the infinite-size limit. 
These findings are consistent with a theoretical study by Firpo \cite{firpo}
 based on a Riemannian approach, which predicts finite $\lambda_1$ values below the transition.
However, recently, Manos and Ruffo \cite{tanos}
 claimed that the $N^{-1/3}$ law in the homogeneous phase
 applies also at low energies,
 specifically for $U < 0.2$, while in the range of $0.2 < U < U_c$
 they are unable to decide whether the
maximal LE vanishes or stays finite.
Finally, a recent statistical-mechanical treatment
 suggests that the Lyapunov spectrum should
 always converge to zero, but cannot exclude the existence of an
 anomalous subextensive component of strictly positive LEs \cite{kurchan}.



In this paper we revisit this issue of the existence and nature of chaos in the HMF model.
The main part of this paper is a study of the largest
LE which is split into two parts: (i) the analysis of
finite-time LEs of a single oscillator in the fluctuating potential
 under the assumption of a negligible coupling in tangent space;
 (ii) a careful investigation of the effect of the tangent-space coupling.
The former analysis is justified by the empirical observation that the
first Lyapunov vector is localized and
 the fact that the influence of a given oscillator on the
 self-consistent mean field, which we call simply the coupling strength,
 decreases as $1/N$. 
We conclude that the single-oscillator LE
 (i.e., the mean of finite-time LEs) cannot be
larger than $1/\ln N$, but we also observe that its fluctuations stay finite in
the thermodynamic limit. These results are related to the existence of a
homoclinic cycle connecting the top of the effective potential with itself.
The following analysis of the coupling in tangent-space
reveals that even if it is very small it induces a finite increase in the LE,
 which is proportional to the fluctuations of the single-oscillator LE. 
This phenomenon, that we
call ``coupling pressure'', is a manifestation of a strong sensitivity to
coupling which generally arises in ensembles of identical weakly-coupled
oscillators. It was first uncovered in two coupled identical ocillators,
where it was shown that the maximum  LE increases with coupling
strength $\varepsilon$  by an amount of order $1/|\ln \varepsilon|$~ \cite{daido}.
The same effect was later found in
higher dimensional systems \cite{LPR,LPRT,CP,piko}. 
In the context of globally-coupled systems,
 the coupling pressure is so drastic that it survives even though the
coupling strength vanishes in the thermodynamic limit. 
We provide a quantitative
explanation of the effect, by mapping the tangent-space evolution onto a
stochastic model of sporadically-coupled diffusing particles
 (see Ref.~\cite{short}
for a preliminary discussion). In the HMF model, the effect of the coupling
pressure is particularly important, since it increases the value of the largest
LE from zero to a finite number, i.e., it induces an instability in
a model that would, otherwise, be non-chaotic. Altogether, we can
summarize our results by stating that the infinite-size and the infinite-time
limits do not commute: taking first the thermodynamic limit, the evidence of
dynamical instabilities would be lost. An indirect confirmation of the
theoretical approach comes from the localization of the Lyapunov vector, that
is confirmed by our numerical simulations.

Our investigation of the largest LE in the ordered phase of the simple HMF model
is completed by the study of a number of related points:
first, we numerically study the largest LE in the vicinity of the critical energy value $U_{\rm c}$. We find that
$\lambda_1$ goes to zero for $U\to U_{\rm c}$ from below and account for the observed
scaling behavior.

Next we address the problem of the shape of the entire Lyapunov spectrum.
We find that several exponents (in addition to the largest) stay positive, but their
number is non-extensive (i.e., it grows slower than linearly with $N$). The
results are consistent with the theory in Ref.~\cite{kurchan}
 where it was predicted
that ``with measure one" the spectrum should be equal to zero.

Finally, we discuss a 2D generalization
 of the HMF model \cite{art,AntoniTorcini},
to test the general validity of our theoretical and numerical
findings. 
In the 2D model, each oscillator is composed of four variables
 and thus can be chaotic without taking
 the coupling pressure into account.
We find nevertheless the same size-dependence of the largest LE
 as in the standard HMF model.
We also investigate the full Lyapunov spectrum in this case
 and argue to what extent the argument developed
 for the standard HMF can be extended here.

This paper is organized as follows:
The HMF model is introduced in Sec.~\ref{HMF}, together with a
careful discussion of its equilibrium properties. This is
necessary to collect proper information on finite-size effects that is
crucial for a correct development of our theoretical arguments. Section
\ref{Lyapunov} is devoted to a critical discussion of the numerical results
for different system sizes and different energy values. There, we illustrate
some of the issues that hindered the interpretation of the numerical results.
Section~\ref{single-oscillator}
 is devoted to a detailed characterization of the
evolution in the tangent space of a single particle in a self-consistent
mean field. In particular, we introduce a finite-time LE and
discuss its dependence on the energy and the number of particles.
The effect of the coupling is discussed in Sec.~\ref{SecToy}, where we first
introduce a simplified model and test the correctness of our solution. The
application
to the HMF model is analyzed in the second part of the section. The scaling of
the largest LE at the transition energy $U_{\rm c}$ is derived in
Sec.~\ref{sec:critical} and compared with numerics. The structure of the
Lyapunov spectra is analyzed in Sec.~\ref{sec:full}. In
Sec.~\ref{sec:generalized} we deal with the 2D generalization of the HMF model.
A brief summary of the results and a discussion of open questions are
finally reported in Sec.~\ref{sec:conclusions}.

\section{The Hamiltonian mean field model: equilibrium results}
\label{HMF}

The HMF model was derived from a one-dimensional self-gravitating model,
by truncating the Fourier expansion of the gravitational potential to its first
term \cite{hmf}. 
It consists of $N$ unit-mass particles that move on a circle under
their mutual attraction.  The dynamics of the $N$ particles is ruled by the
Hamiltonian \cite{noteJ}
\be
H = K + V \equiv \sum_{i=1}^N \frac{p_i^2}{2} +
\frac{1}{2N} \sum_{i,j=1}^N \left[1-\cos(\theta_i-\theta_j)\right] \, ,
\label{hmf1}
\ee
where $\theta_i$ and $p_i$ denote particle positions (angles) and velocities.
The resulting equations of motion write
\bey
&&\dot \theta_i = p_i,\nonumber \\
&&\dot p_i = \frac{1}{N} \sum_j \sin(\theta_j-\theta_i) = M \sin(\phi-\theta_i),
\label{model1}
\eey
where $M$ is the magnetization and $\phi$ the associated phase, defined by
\be
 M {\rm e}^{i\phi}  = \frac{1}{N} \sum_j {\rm e}^{i\theta_j}.
\label{model2}
\ee
$M$ measures the degree of clusterization and plays the role of an order
parameter \cite{kuramoto}. Depending on the energy density $U=H/N$, the
system can show two different thermodynamic phases, separated by a
second-order transition: (i) the clustered, ordered phase, characterized by a finite
magnetization (for $U<U_{\rm c}=3/4$) ; (ii) the homogeneous phase, characterized
by a vanishing magnetization (for $U>U_{\rm c}$). 
In the following, we limit ourselves to the clustered phase
$U \le U_{\rm c}$ ($T \le T_c$).


All the reported simulations have been performed within the microcanonical
ensemble by implementing symplectic integration schemes, typically a 4-th order
McLahlan-Atela algorithm \cite{mca} with integration time step $dt=0.05$ or 0.1.
This choice ensures an energy conservation with a relative precision of the
order of $10^{-10}$ to $10^{-11}$.

Initial phases and
momenta have been typically drawn from the invariant equilibrium
distribution discussed in the next subsection. We have also occasionally
compared the results with those obtained for different choices of initial
conditions, namely: (i) zero phases and a Gaussian distribution of the momenta;
(ii) a uniform distribution of the phases and a Gaussian distribution of the
momenta. A transient (typically $5 \times 10^3 N$ time
units) has been discarded, before starting
the computation of any equilibrium quantity.
Finally the typical transient time for the evolution in tangent space lies
between $4 \times 10^5$ and $4 \times 10^6$ time units, while the typical
integration time lies between $4 \times 10^6$ and $ 10^7$ time units.

\subsection{Equilibrium distribution of single oscillator energy}

From the point of view of a single oscillator,
 the evolution equation in Eq.~\eqref{model1},  at finite system sizes,
is equivalent to that of a pendulum in a noisy environment, the noise
being the result of the statistical fluctuations of the magnetization.
In the thermodynamic limit, $M$ and $\phi$ are strictly constant and, as a
result, the single oscillator energy 
\be
h_i = \frac{p_i^2}{2} + M[1-\cos(\theta_i-\phi)] \equiv k_i + v_i
\label{H1}
\ee
is strictly conserved. Notice that, in Eq.~(\ref{H1}), the (arbitrary) zero
level of the potential energy $v_i$ is shifted by $v_M=M-1$ in
order that the ground-state energy of the single oscillator is always zero
 for any value of the magnetization $M$.
This is the convention adopted throughout the 
paper \cite{Note1}.
Note also that the total potential
energy is not given by the sum of the single potential terms, but it is equal
to half of it, namely
\begin{equation}
V = \frac{1}{2}
\sum_{i=1}^N (v_i-v_M),
\end{equation}
the reason being that each term would otherwise be counted twice.

\begin{figure}
\includegraphics[draft=false,clip=true,width=\columnwidth]{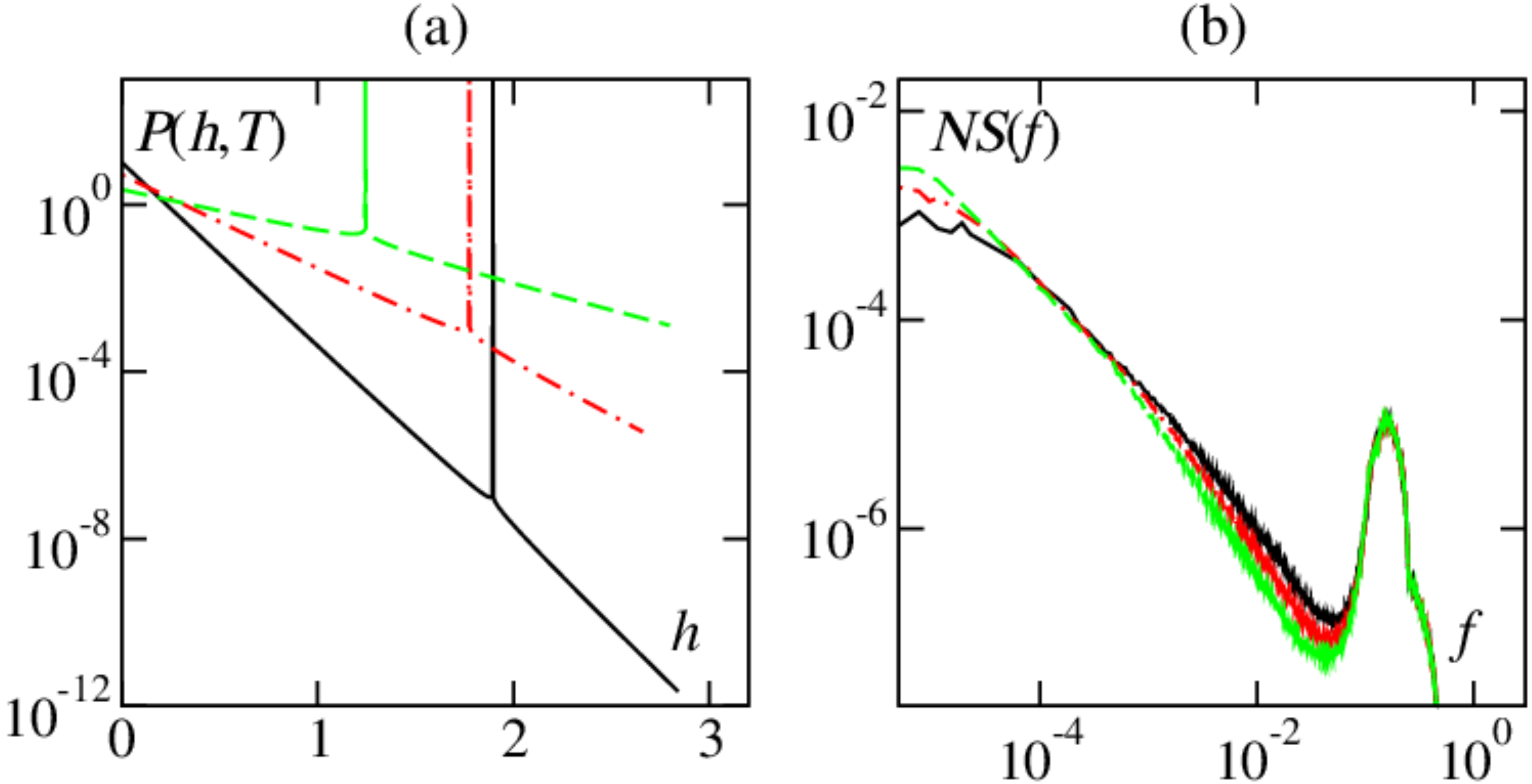}
\caption{(color online). (a) Equilibrium energy distribution [Eq.~\pref{singleparticleP}]
 for three different values of 
the internal energy: $U = 0.1$, $U=0.2$, and $U=0.5$ (from bottom to top).
The value of $T$ in Eq.~\pref{singleparticleP} is computed by Eq.~\pref{energytemp}.
Recall the shifted energy axis, Eq.~(\ref{H1}), for $h$.
(b) Power spectrum $S(f)$, multiplied by the size $N$,
 for $U=0.5$ and three different
system sizes, $N=250$ (solid black line), 500 (dotted dashed red line),
and 1000 (dashed green line).} 
\label{fig1}
\end{figure}

The equilibrium distribution of the single-oscillator energies is
\begin{align}
P(h,T)  = &\int_0^\infty dp \int_0^\pi d\theta \nonumber \\
 &\times \delta \left[h- \frac{p^2}{2} - M(1 - \cos \theta) \right] Q(p,\theta,T),
\end{align}
where 
$Q(p,\theta,T)$ is the Gibbs-Boltzmann distribution
\begin{equation}
Q(p,\theta,T) = C \exp \left[ -\frac{p^2}{2T}-\frac{M}{T}(1 - \cos \theta) \right],
\end{equation}
 with a suitable normalization constant $C$,
 the unit Boltzmann constant, and the temperature $T$ given
 by \cite{hmf,rev1,rev2}
\be
U=\frac{T}{2}+\frac{1}{2}(1-M^2).  \label{energytemp}
\ee
As a result, one finds that
\begin{equation}
P(h,T)  = 
\frac{C}{\sqrt{2M}} \int_0^{y_0} \frac{{\rm e}^{-h/T}}{\sqrt{y(h/M-y)(2-y)}} dy, 
\label{singleparticleP}
\end{equation}
where $y_0=h/M$ if $h/M<2$ and $y_0=2$ otherwise. The integrand has two
(integrable) square-root singularities at both ends of the integration interval
for all energy values, except for $h=2M$, in which case the singularity is
hyperbolic, and this indicates a logarithmic divergence of the integral.
Note that the divergence lies at the separatrix $e_{\rm s}$ of the single-particle
effective Hamiltonian (\ref{H1}). This equilibrium energy distribution
 \pref{singleparticleP} is plotted in Fig.~\ref{fig1}(a)
 for three
different energy densities ($U=0.15$, 0.2 and 0.5).

\subsection{Magnetization}
\label{subsec:magn}

In this section we discuss the behavior of the magnetization in the clustered
phase and at the critical energy $U_{\rm c}$ for finite $N$.  For large but finite
$N$, the magnetization is affected by statistical fluctuations and, as a result,
the oscillator energies diffuse, albeit very slowly. Simple statistical arguments
suggest that the absolute value of the magnetization $M(t)$ fluctuates around
its mean field value  with an amplitude that should scale as $1/\sqrt{N}$. 
We numerically checked this conjecture by
measuring the power spectrum $S(f)$ of the magnetization $M(t)$
 for different system sizes. In
Fig.~\ref{fig1}(b), we see that most of the power is concentrated in a broad peak
around $f =0.16$ (for $U=0.5$) and the peak power scales as $1/N$,
 in agreement with the $1/\sqrt{N}$ amplitude of the fluctuations.
We also verified that the power
contained in the low-frequency peak increases upon increasing the system
size. 
The peak location should be related to some
characteristic timescale of the dynamics, though we have no precise hints about its origin.

\begin{figure}
\includegraphics[draft=false,clip=true,width=\columnwidth]{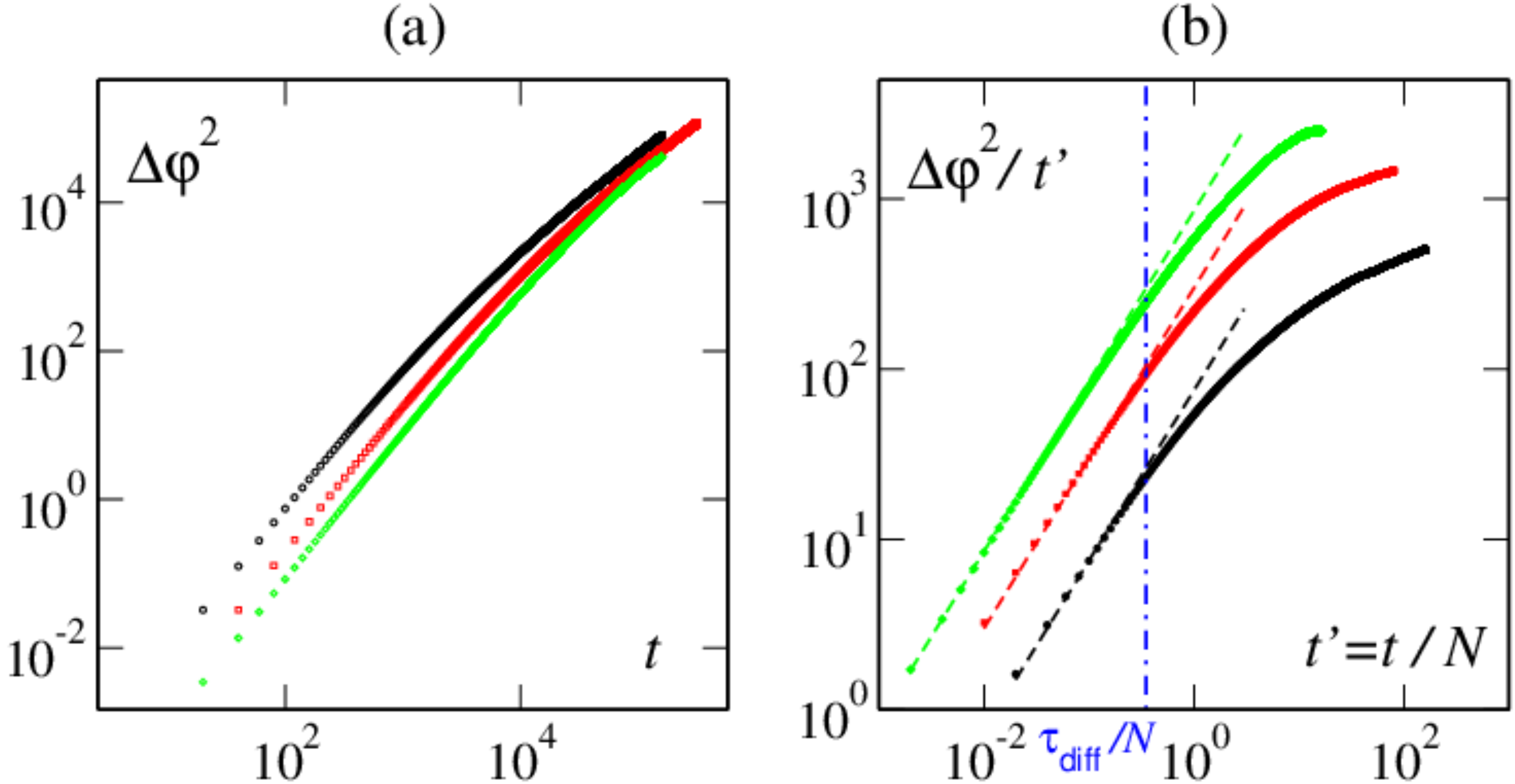}
\caption{(color online). (a) Mean-square displacement of the global magnetization phase
$\Delta \phi^2=\langle (\phi(t)-\phi(0))^2 \rangle$ vs. time. Mean-square displacements 
have been computed by averaging over time span of order $1 \times 10^7$ with sliding
windows of duration $160000 - 320000$ for three different system sizes: namely, $N=1000$ (black dots),
$N=4000$ (red squares) and $N=10000$ (green diamonds). 
(b) Same quantities as in (a) but rescaled to better highlight the crossover from ballistic to diffusive behavior. 
Time has been rescaled by system size, $t'=t/N$,
and mean-square displacements by the rescaled time, $\Delta \phi^2 \to \Delta \phi^2 / t'$. 
The dashed lines mark the linear ballistic growth, and the vertical dot dashed
(blue) line 
highlights the beginning of deviations from linear growth at the rescaled crossover time $\tau_{\rm diff}/N$.} 
\label{fig:taudiff}
\end{figure}

In contrast, the motion of the global phase is determined by the oscillators
with single-particle energies $h$ larger than the separatrix energy $e_{\rm s}$
\cite{hmf}. 
They continue rotating
 either clockwise or counterclockwise according to their momentum $p$. 
The slow energy diffusion of individual oscillators implies that
 those with low energies wander and can reach $h>e_{\rm s}$,
``randomly" picking the rotation direction, and stay in this high-energy state
for some time until they eventually go back to the ``bounded'' state
 with $h<e_{\rm s}$. The
numbers of particles rotating clockwise and counterclockwise are on average
equal to one another, but, because of statistical fluctuations, the
instantaneous fractions of the populations
 typically differ by a quantity of order $1/\sqrt{N}$.
Because of momentum conservation, the phase
$\phi_N$ \cite{Note2} of the global
magnetization exhibits a net drift
 with an average angular velocity $\omega
\sim 1/\sqrt{N}$ \cite{hmf}, which has also been verified numerically (not shown). 
Over long time scales, 
the sign of the velocity changes since the fluctuations
will invert the predominance of clockwise/counterclockwise rotating particles. 
As a consequence, we expect the global phase to exhibit a crossover from
drifting to diffusive motion over a crossover timescale $\tau_{\rm diff}$. Numerical simulations
(see Fig.~\ref{fig:taudiff}) clearly show that the crossover time diverges in the thermodynamic 
limit as $\tau_{\rm diff}\sim N$.

Finally, we discuss the equilibrium behavior in proximity of the critical
energy. In the thermodynamic limit, the magnetization $M$ obeys the usual
mean-field behavior, i.e., $M \sim |U - U_{\rm c}|^{1/2}$ for $U < U_{\rm c}$
\cite{rev1,rev2}. Determining $M$ for finite $N$ is, however, a delicate problem which
requires a careful treatment of finite-size fluctuations
\cite{Daido-JSP1990, Pikovsky_Ruffo-PRE1999, Hong_etal-PRL2007}.
The correct solution can be found by taking into account the law of large
numbers in the self-consistency equation of the mean field argument. The
magnetization can be expressed within the canonical ensemble formulation as
\begin{equation}
 M + \delta M = \left | \frac{1}{Z} \int_{-\infty}^\infty \rd p \int_0^{2\pi} \rd
\theta \e^{i\theta} \e^{-\mathcal{H}/T} \right |,  \label{eq:crit1}
\end{equation}
where $\mathcal{H} = p^2/2 + 1 - M \cos\theta$ (here the absolute scale
is chosen for the energy),
$Z = \int_{-\infty}^\infty \rd p \int_0^{2\pi} \rd \theta \e^{-\mathcal{H}/T}$,
and $| \cdot |$ denotes the modulus of the complex number. 
The second term in the l.h.s. represents an
unavoidable finite-$N$ correction that we assume to be in the order of
$\delta M \sim \mathcal{O}(M/ \sqrt{N})$.
A straightforward calculation (see also Ref.~\cite{hmf}) leads to the self-consistency
equation
\begin{equation}
 M = \frac{I_1(M/T)}{I_0(M/T)} - \delta M,  \label{eq:crit2}
\end{equation}
where $I_n(z)$ is the first-kind modified Bessel function of order $n$.
Near the critical point, Eq.~\eqref{eq:crit2} can be expanded
for small $M$, yielding
\begin{equation}
 M \simeq \frac{1}{2}
\frac{M}{T} \[ 1-\frac{1}{8}\left(\frac{M}{T}\right)^2 \] - \delta M \quad.
\label{eq:crit3}
\end{equation}
In the infinite size limit, $\delta M = 0$, and this gives the expected
mean-field result \cite{rev1,rev2}, $M \sim (T_c - T)^{1/2}$ for
$T < T_{\rm c} = 1/2$. 
For finite sizes $N$, Eq.~\eqref{eq:crit3} predicts
the following scaling,
\begin{equation}
 M(U, N) \sim N^{-\beta/\nu}
F \left((T_{\rm c}-T) N^{1/\nu}\right),~~~~ \text{for $T < T_{\rm c}$,}
\label{eq:crit4}
\end{equation}
with $\beta = 1/2$, $\nu = 2$, and a scaling function $F(z)$.
\begin{figure}[t]
  \includegraphics[width=\hsize,clip]{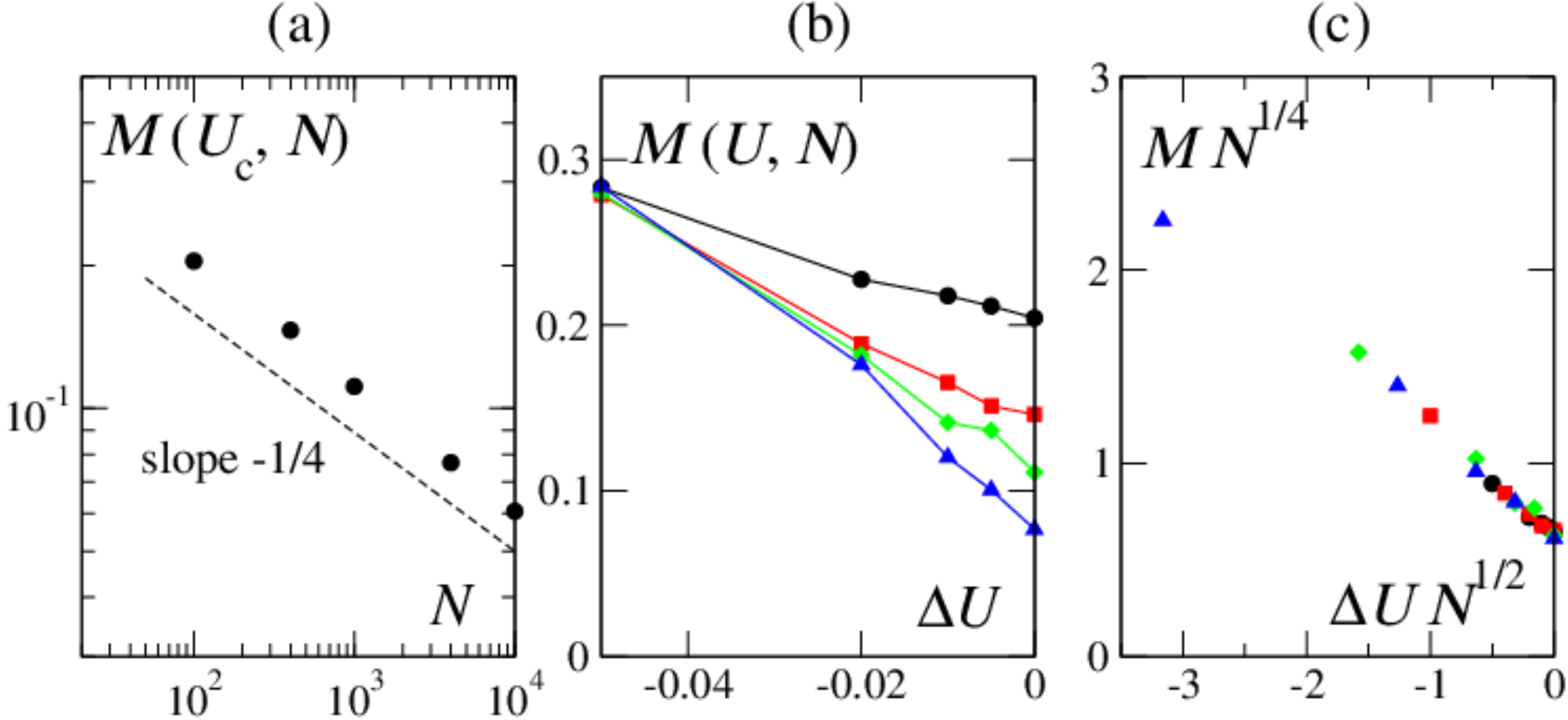}
\caption{(color online). Critical scaling of the magnetization $M$.
(a) Size dependence of the magnetization at the critical point, $M(U_{\rm c}, N)$.
(b) Magnetization $M(U,N)$ as a function of the energy difference
$\Delta U=U-U_{\rm c}$ for $N = 100, 400, 1000$, and 4000 (from top to bottom).
(c) Same data as the panel (b) with rescaled axes.}
  \label{fig:crit1}
\end{figure}
Using the energy-temperature relation \eqref{energytemp} \cite{Note3},
we can rewrite Eq.\ \eqref{eq:crit4} as
\begin{equation}
 M(U, N) \sim N^{-\beta/\nu} G \left( (U_{\rm c}-U)
N^{1/{\nu}} \right),~~~~ \text{for $U < U_{\rm c}$,}  \label{eq:crit5}
\end{equation}
with another scaling function $G(z)$.
This expression accounts for the critical decay of the magnetization
$M(U_{\rm c}, N) \sim N^{-1/4}$ found in Fig.\ \ref{fig:crit1}(a).
Equation~\eqref{eq:crit5} can be further checked by rescaling the magnetization
$M(U,N)$ off criticality; plotting $MN^{\beta/\nu}$ against
$(U-U_{\rm c}) N^{1/\nu}$, we confirm that the data shown in
Fig.~\ref{fig:crit1}(b) collapse reasonably well onto a single curve,
$G(z)$ [Fig.~\ref{fig:crit1}(c)]. 
It is worth noticing that the observed finite-size scaling
 $M(U_{\rm c}, N) \sim N^{-1/4}$ for the magnetization was
 reported for the first time for the mean field
version of the Ising and Heisenberg model in Ref.~\cite{kittel}.
The obtained value of the critical exponent ${\nu} = 2$
 is the one found for dissipative noisy phase oscillators
\cite{Pikovsky_Ruffo-PRE1999}, but is different from that of the Kuramoto
model, i.e. deterministic phase oscillators with random frequencies,
${\nu} = 5/2$ \cite{Hong_etal-PRL2007}. 
The value ${\nu} = 2$ is also the one
expected from a simple dimensional analysis, ${\nu} = \nu_{\rm MF} \,d_{\rm c}$,
where $\nu_{\rm MF}$ is the usual correlation-length exponent in the mean-field
limit and $d_{\rm c}$ is the upper critical dimension \cite{Botet_etal-PRL1982}.
In our case, $d_c=4$ and $\nu_{\rm MF}=1/2$, which further confirms the analogy
of the HMF model with the mean field XY Heisenberg model \cite{hmf}.

\section{Lyapunov characterization of the dynamics}
\label{Lyapunov}

In order to characterize the dynamics of the system, we estimate the
LEs by following the dynamical evolution in tangent space 
of a vector ${\bf v}=\{\delta \theta_i, \delta p_i\}_{i=1,\ldots,N}$ of
infinitesimal perturbations,
\bey
&&\delta \dot \theta_i = \delta p_i,
\label{eq:tang1}\\
&&\delta \dot {p}_i = -M \cos(\phi-\theta_i) \delta \theta_i
+\frac{1}{N} \sum_j \cos(\theta_j-\theta_i) \delta \theta_j,
\nonumber 
\eey
and by orthonormalizing the resulting vectors at proper times \cite{Benettin}.  
In order to probe large system sizes up to $N=10^6$, we have implemented a highly 
parallelized version of the Gram-Schmidt algorithm.

Most of the numerical simulations have been performed for $U=0.7$ (which
corresponds to a magnetization $M \approx 0.281$ and a temperature
$T \approx 0.479$, measured at $N \geq 10^5$)
and $U=0.5$ ($M \approx 0.621$ and $T \approx 0.386$).
Both parameter values are sufficiently away from the critical point $U_{\rm c}$
as well as from the zero-temperature limit;
otherwise the asymptotic behavior of the
Lyapunov exponent would be masked by severe finite-size effects (see below).

\begin{figure}[t]
 \begin{center}
\includegraphics[width=\hsize,clip]{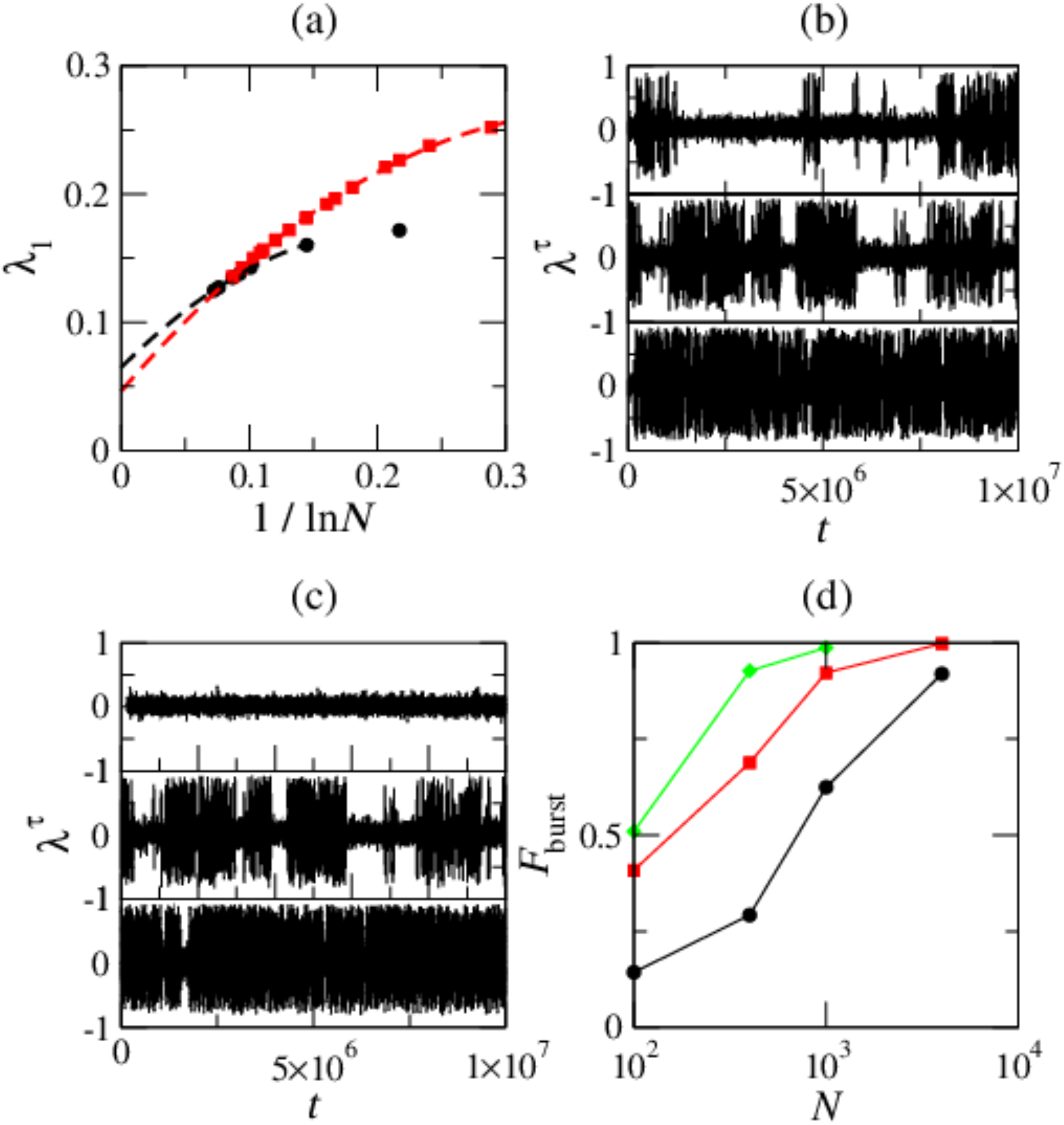}
\caption{(color online). Largest LE $\lambda_1$ of the HMF model. (a)
$\lambda_1$ versus the system size $N$ for $U = 0.5$ (black circles) and $U=0.7$
(red squares). The dashed lines show the quadratic extrapolations to the asymptotic values.  
(b,c) Time series of the finite time
exponent $\lambda^\tau$ for $N = 100, 400, 1000$ with $U = 0.25$
 (panel (b) from top to bottom) and
for $U = 0.20, 0.25, 0.30$ with $N = 400$ (panel (c) from top to bottom).
The laminar and burst states have different values of the time-averaged LE as shown in Fig.~\ref{fig5}(c) below. (d) Time fraction in the burst state $F_{\rm burst}$ as a function of the size $N$ 
for $U = 0.25$ (black circles), $0.30$ (red squares), and $0.35$ 
(green diamonds).}
  \label{fig2}
 \end{center}
\end{figure}

According to a theoretical argument briefly sketched in
Ref.~\cite{short}, where we showed that, in globally-coupled
dissipative systems, the leading finite-size corrections to the
asymptotic value of the largest LE are polynomial in $1/\ln N$, we
find it convenient to investigate the finite-size dependence of $\lambda_1$ 
by plotting it as a function of $1/\ln N$.
The data reported in Fig.~\ref{fig2}(a) are indeed consistent with
 a logarithmic dependence,
\be
\lambda_1 = \lambda_{\infty} + \frac{c}{\ln N} + \mathcal{O}\left( \frac{1}{\ln^2 N} \right),
\ee
for large $N$, especially for $U=0.7$ (red squares).
Deviations from the $1/\ln N$ behavior are visible
 for small $N$, but taking a quadratic correction into account is sufficient
 to describe perfectly all system sizes studied (Fig.~\ref{fig2}(a), red dashed line).
This quadratic correction is stronger for $U=0.5$,
 which is an incipient evidence of the convergence problems
 that arise at small energies (see below).
Because of this slow but significant size-dependence,
 the extrapolated values of the maximum LE in the
thermodynamic limit are smaller than the typical values reported in the
literature (see,  e.g., Ref.~\cite{rev1}),
 but are neverthless clearly different from
zero: $\lambda_{\infty}=0.056(6)$ at $U=0.5$ and
$\lambda_{\infty}=0.046(3)$ at $U=0.7$, 
both obtained by using the quadratic ansatz described above.

As briefly mentioned in the introduction,
 a number of numerical studies have reported
 contradicting conclusions about the largest LE in the clustered phase:
 while most of them claimed, qualitatively,
 no or weak size-dependence and thus strictly positive asymptotic values
 of the maximal LE \cite{rev1,latora98}, Manos and Ruffo \cite{tanos}
 reported a power law decay $\lambda \sim N^{-1/3}$,
 although their simulations were performed for substantially lower
energy densities ($U=0.1$). Our own numerical simulations (not shown) performed
at $U=0.1$ indeed confirm the power-law decay at least up to $N=10^6$. In order
to shed some light on the possible existence of two qualitatively different
phases, we scanned intermediate energy levels in the interval $U \in (0.1,0.5)$.
These simulations (see below) revealed the presence
 of strong intermittent behavior,
 which make practically impossible to determine
 a reliable value of the LE, in particular for large system sizes.
The phenomenon is better illustrated by studying the
finite time LE
\be
\lambda^\tau(t)=\frac{1}{\tau} \ln \frac{||{\bf v}(t)||}{||{\bf v}(t-\tau)||}\,,
\ee 
fixing $\tau=2$ \cite{Note4}.
Fig.~\ref{fig2}(b,c) reveals irregular jumps of $\lambda^\tau(t)$ between
two clearly different states: (i) a {\it laminar} one, where
$\lambda^\tau(t)$ stays near zero; (ii) {\it bursts}, where
$\lambda^\tau(t)$ fluctuates much more strongly. Upon comparing
simulations performed for different sizes [Fig.~\ref{fig2}(b)]
and different energy densities [Fig.~\ref{fig2}(c)],
 we see that the frequency of the
bursts grows both with $N$ and $U$. This is quantitatively shown in
Fig.~\ref{fig2}(d).

In order to understand the origin of this intermittent behavior, we analyze
the structure of the (first) Lyapunov vector.
The Lyapunov vector is the quantity associated with each LE
 and indicates the direction of the infinitesimal perturbations
 growing at the rate of the corresponding LE.
It is defined as a function of the phase-space point
 and turns out to be a useful tool to characterize
 statistical properties of large dynamical systems
 \cite{CLV,collective,spurious}.
Here, it is convenient to introduce the squared amplitude
 of the normalized vector components for each oscillator,
\be
A_i = \delta \theta_i^2 +\delta p_i^2 \, ,
\label{amplitudes}
\ee
and to consider its time average $\langle A_i \rangle$ (here and in
the following, angular brackets denote time averages), to have a statistically
reliable quantity.

\begin{figure}
\includegraphics[draft=false,clip=true,width=\hsize]{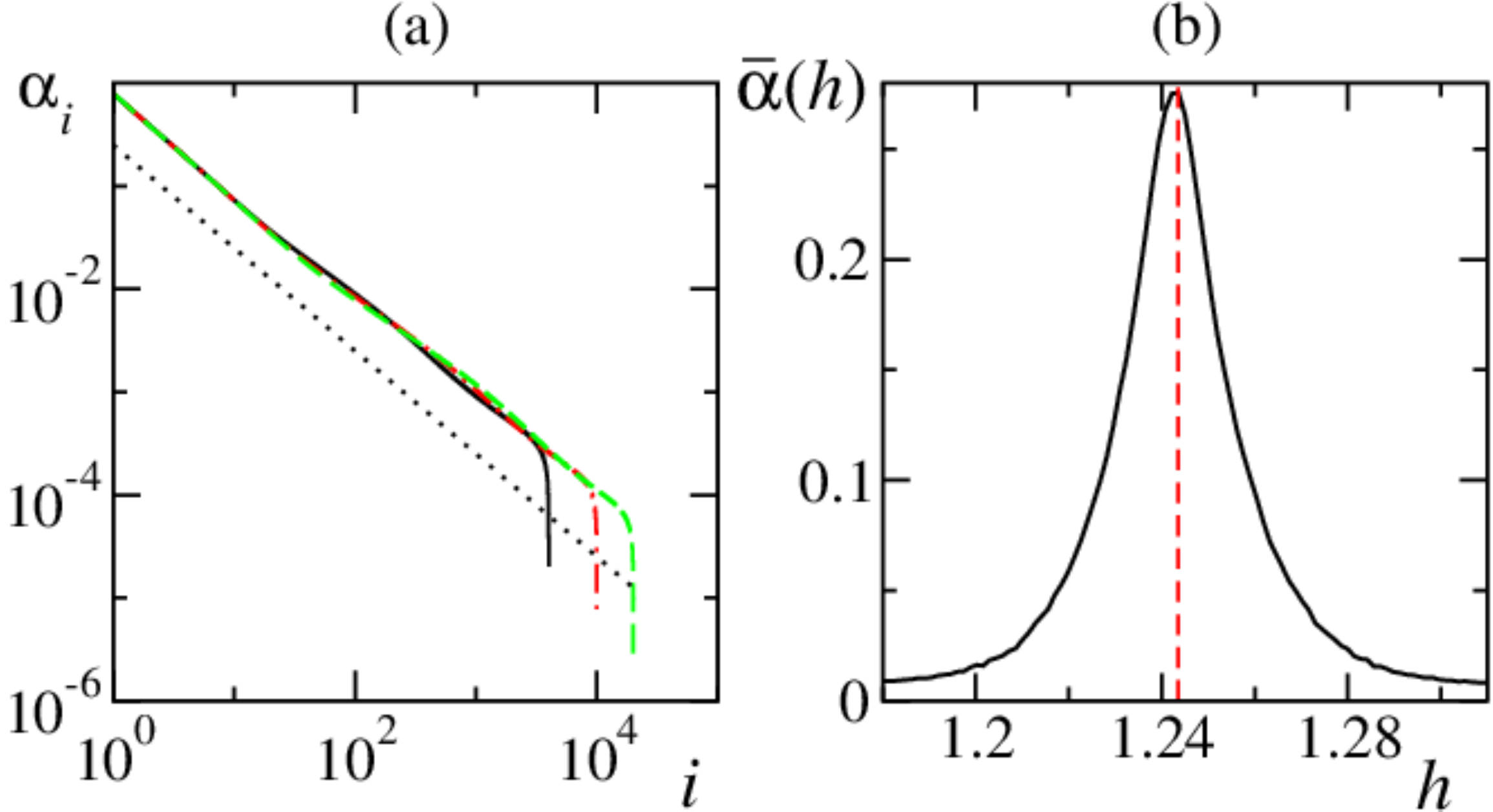}
\caption{(color online). Localization of the Lyapunov vector. (a) $\alpha_i$ vs $i$ for $N=4000$ (black solid line),  $N=10000$ (red dotted dashed line), and  $N=20000$ (green dashed line) at $U=0.5$.
The dotted black line marks a decay as $1/i$. (b) Average amplitude of the vector components as a function of the corresponding single-oscillator energy,
 $\overline\alpha(h)$, in a
system of $N=10^5$ oscillators at $U=0.5$. The vertical dashed red line marks
the single oscillator separatrix energy $e_{\rm s} = 2 M$.
}
\label{fig4}
\end{figure}

In homogeneous globally-coupled systems, any ordering of the oscillators is
equally meaningful, as they are equivalent to one another. In Fig.~\ref{fig4}(a),
we plot the amplitude $\alpha_i = \sqrt{\langle A_i \rangle}$ versus
its rank (i.e., we arrange the oscillators according to $\alpha_i$ in
decreasing order)
for $N=4000$, $10000$, and $20000$. 
Our data show that
the Lyapunov vector is approximately localized as $1/i$ (as indicated by the dotted black
line), that is, the perturbation is concentrated in a few components. In
Fig.~\ref{fig4}(b),
 the data is organized in a different way. The oscillators are
grouped according to their energy and the perturbation amplitude is averaged
over all oscillators in the interval $[h,h+dh]$, to obtain $\overline
\alpha(h)$. The results reported in Fig.~\ref{fig4}(b) indicate that the
vector component is substantially larger when the energy of the corresponding
oscillator is close to that of the separatrix.

In order to further clarify
the relationship between localization and energy, it is instructive
to monitor the instantaneous degree of localization of the Lyapunov vector,
 by estimating the inverse participation ratio \cite{Mirlin-PhysRep2000}
\be
Y_2=\frac{\sum_i A_i^2}{(\sum_i A_i)^2}.
 \label{participation}
\ee
By construction, $1/N \leq Y_2 \leq 1$. The larger is $Y_2$, the more the vector is
localized; $Y_2=1$ denotes complete localization on a single oscillator,
while $Y_2 = 1/N$ indicates a completely delocalized vector with equal components.

\begin{figure}
\includegraphics[draft=false,clip=true,width=\hsize]{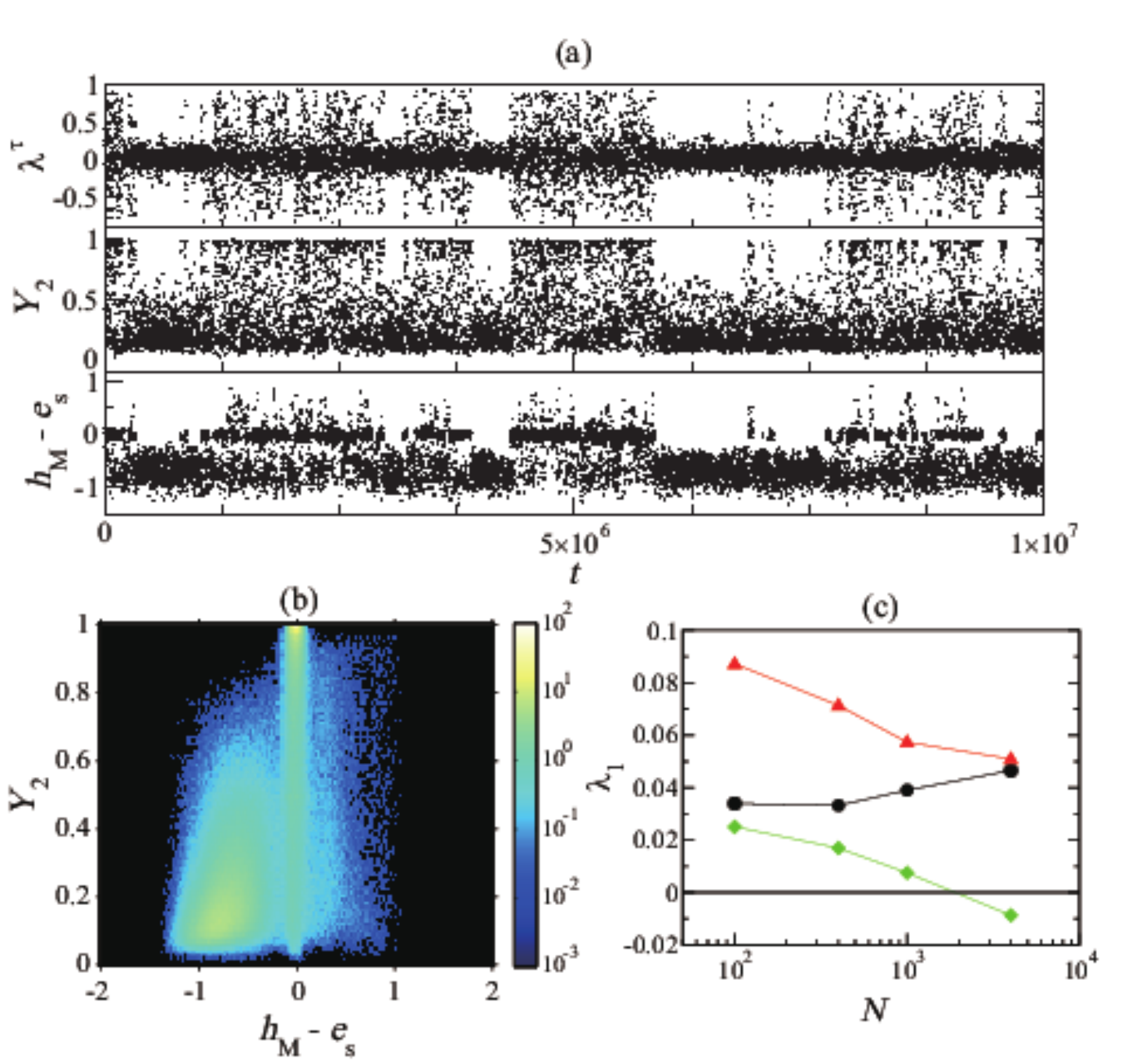}
\caption{
(color online). Intermittency for $U = 0.25$ and $N = 400$. (a) Time series of the
finite-time LE $\lambda^\tau$ (time window $\tau=2$),
the instantaneous value $Y_{2}$ of the inverse participation
ratio, and the energy difference $h_{\rm M} - e_{\rm s}$ between the
dominating oscillator (the one with the largest amplitude $A_i$ of the Lyapunov vector component) and the separatrix. (b) Density plot with respect
to $h_{\rm M} - e_{\rm s}$ and $Y_{2}$. The color code shows the
frequency in the logarithmic scale. The black indicates null density.
Two peaks corresponding to the burst
and the non-burst states are clearly visible. (c) Lyapunov exponent $\lambda_1$ (black circles) versus $N$ for $U=0.25$. The
conditioned Lyapunov exponents (see text) are also shown for the burst state (red triangles) and for the non-burst state (green diamonds).
}   
\label{fig5}
\end{figure}

In Fig.~\ref{fig5}(a), we plot the time evolution of the finite-time LE
 $\lambda^\tau$, of the inverse participation ratio $Y_2$, and of the
energy $h_M$ of the oscillator with the largest amplitude $A_i$
 in the Lyapunov vector components.
The data refer to a small system ($N=400$) with energy density $U=0.25$. 
The three temporal
traces reveal a strong correlation
 between the occurrence of the bursts in the finite-time LE, a
stronger localization and the closeness of the energy
 to that of the separatrix.
A more quantitative characterization of the connection between the inverse
participation ratio and the energy is presented in Fig.~\ref{fig5}(b),
 where the color code indicates the probability to observe a given pair
 of values $(Y_2,h_M-e_{\rm s})$.
Altogether, the data plotted in Fig.~\ref{fig5}(a,b) confirm the intermittency
 between the two distinct states: (i) the laminar state is
characterized by a less fluctuating finite-time LE,
 a weak localization of the Lyapunov vector, and
 single-oscillator energies far from the separatrix; 
(ii) the burst state by large fluctuations of the finite-time LE
 and a strong localization of the Lyapunov vector
 around an oscillator lying very close to the separatrix.

In Fig.~\ref{fig5}(c)
 we compare the value of the true, time-averaged LE (circles)
with the averages restricted to the bursts (triangles) and the laminar state
(diamonds) for $U=0.25$. We see that upon increasing the system size, the
``burst'' LE tends to converge towards the true LE. This reflects
the fact that the laminar state tends to disappear for $N\to\infty$
 and the ``laminar'' LE remains quite small.

Therefore, we conclude this numerical analysis by noticing that the
observed value of the LE depends strongly on whether there is at least one
oscillator whose energy is sufficiently close to the energy $e_{\rm s}$ of the
separatrix. If, for any reason, no oscillator has an energy $h_i$ close enough
to $e_{\rm s}$, the laminar contribution dominates.
Altogether, our analysis suggests that a truly
asymptotic behavior is observed only if (on average) at least one oscillator
has an energy sufficiently close to that of the separatrix. The minimal number
$N_m$ ensuring this condition can be estimated by imposing
 $N_m P(e_{\rm s},T)\delta h =1$
 with a suitable width $\delta h$ of the energy window. 
It grows quickly with decreasing $U$,
 since, for low energies,
 the energy distribution is approximately exponential, 
 $P(h,T) \approx \exp(-h/T)$ with $T \approx 2U$ [see Eq.~\eqref{energytemp}],
 while the separatrix energy is practically constant, $e_{\rm s} = 2M \approx 2$.
By referring to the
theoretical expression in Eq.~(\ref{singleparticleP})
 and assuming $\delta h \approx 1/\sqrt{N}$
 (see the next section for a justification), we find that $N_m
\approx 10^{13}$, $10^5$, and $10$ for $U=0.1$, 0.2, and 0.5,
 respectively (these are the energy values considered in Fig.~\ref{fig1}). 
It is clear
that for $U=0.1$ there is no hope to reach the asymptotic regime in
numerical simulations with the currently available machines
 and, in particular, that the $\lambda_1 \sim N^{-1/3}$ scaling
 found by Manos and Ruffo \cite{tanos} characterizes only the laminar state,
 which is not the asymptotic state of the system.

Although our numerical results strongly support
 the strictly positive asymptotic value of the maximal LE,
 it is not clear how this behavior is
connected to the presence of oscillators in the vicinity of the separatrix. 
The following two sections are devoted to clarifying this point.

\section{Single oscillator analysis}
\label{single-oscillator}

In this section we analyze the behavior of the Lyapunov exponent and vector,
 by neglecting the coupling
term in the tangent space [i.e., the sum in Eq.~\eqref{eq:tang1}]. This
assumption
 is tantamount to 
studying a single oscillator forced by the
field $M_N(t) {\rm e}^{i\phi_N(t)}$, which is
generated self-consistently by an ensemble of $N$ globally-coupled oscillators.
The evolution is ruled by the
effective Hamiltonian (\ref{H1}) and is thereby described by the equation
\bey
&&\dot \theta = p\nonumber \\
&&\dot p =  M_N \sin(\phi_N-\theta) \, ,
\label{forcedoscillator1}
\eey
which, in the tangent space, becomes
\bey
&&\delta \dot \theta = \delta p\nonumber \\
&&\delta \dot p = -M_N \cos(\phi_N-\theta) \delta \theta\,.
\label{forcedoscillator2}
\eey
with no contribution from the coupling with the other oscillators.
As already noted, the two observables $M_N$ and $\phi_N$ are strictly constant
 in the thermodynamic limit $N \to \infty$. This means that the corresponding
LE is expected to be equal to that of a standard pendulum, i.e., zero.


In the following, we investigate the size-dependence of the maximum LE
 of the single forced oscillator, by introducing an
energy-dependent single-particle LE $\lambda_0(h)$. 
Since a meaningful definition of a LE involves the
infinite-time limit, while energy is conserved only during finite times
 at finite $N$, we introduce sporadic small corrections to prevent
the trajectory from diffusing away from the prefixed energy shell $h$. This is
achieved by rescaling the kinetic energy, or the particle velocity,
each time the trajectory
passes through the point of minimal potential energy,
without adjusting the potential energy.

\begin{figure}
\includegraphics[draft=false,clip=true,width=\hsize]{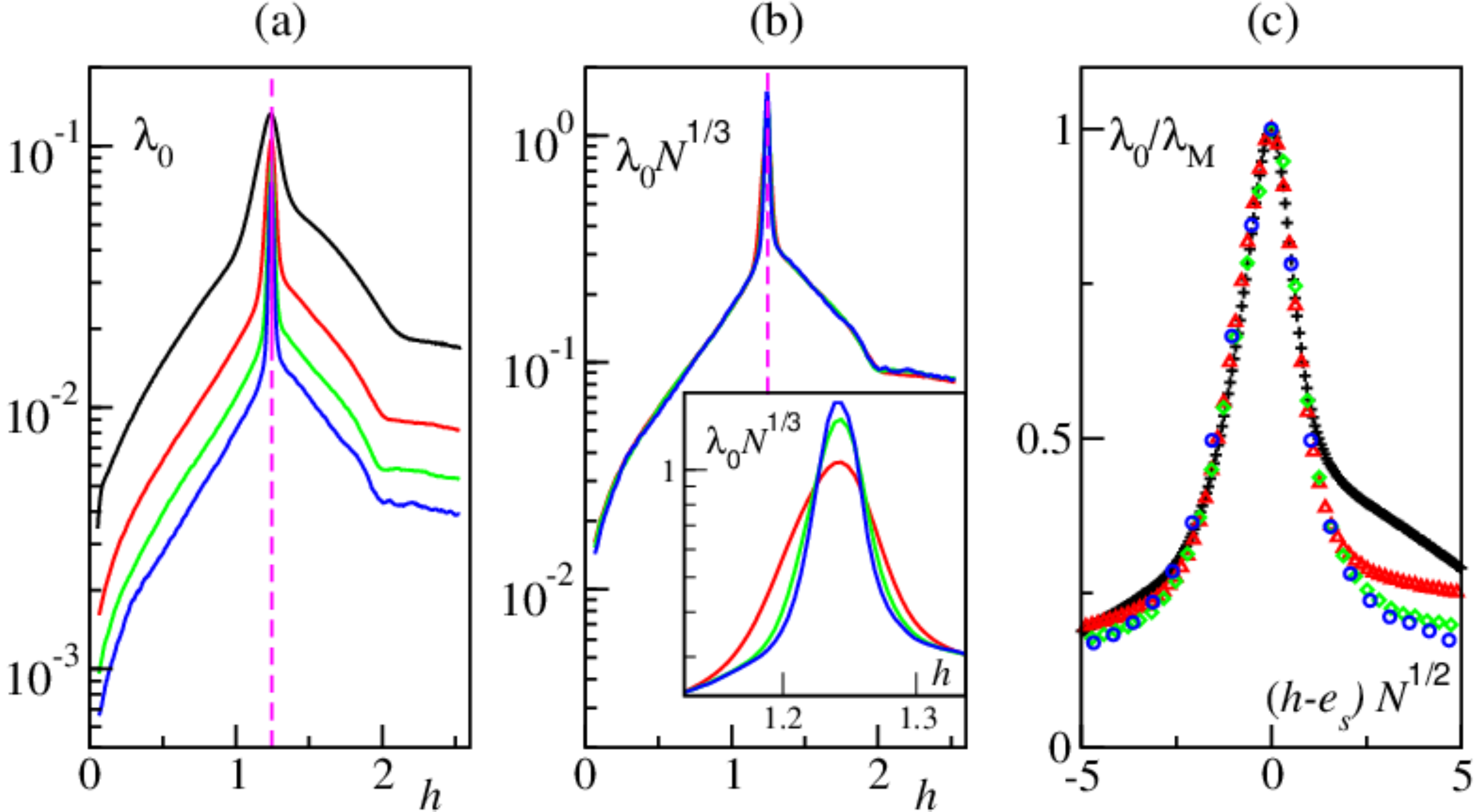}
\caption{(color online). Single-oscillator LE versus the energy of
the single particle, forced by $N$ globally-coupled oscillators at $U=0.5$. In panel (a) the solid curves correspond to $N=100$, $1000$, $4000$, and $10000$ (from top to bottom); the three rescaled curves in panel (b) correspond to $N=1000$, $4000$, and $10000$. The curves overlap except in the peak region, or near the separatrix energy. The vertical dashed lines indicate the energy $e_{\rm s}$ of the separatrix. Panel (c) shows a close-up of the peak at $h=e_{\rm s}$, the LE being scaled by the maximum value $\lambda_{\rm M}$ of each curve and plotted against rescaled single particle energies (crosses, triangles, diamonds, and circles correspond to $N=100$, $1000$, $4000$, and $10000$, respectively).}
\label{fig:fily}
\end{figure}

The numerical results reported in Fig.~\ref{fig:fily} confirm that the LE
 decreases with increasing $N$ as expected,
 but it also displays a strong dependence on the energy. 
The peak is centered at the energy $e_{\rm s}$ of the separatrix.
In panel (b) we see that everywhere except in the peak area, $\lambda_0$ scales
as $N^{-1/3}$. This behavior can be understood by invoking known results for
random symplectic matrices, similarly to the maximal LE of the full system
 in the homogeneous phase. It is known \cite{random} that
 approximating the tangent-space dynamics with a product of independent
 symplectic random matrices with zero-mean disorder of amplitude $\eta$
gives rise to a positive Lyapunov exponent which scales as $\eta^{2/3}$.
In our setup,
the statistical fluctuations of the collective magnetization scale as
$1/\sqrt{N}$ and play the role of the disorder. As a result,
 $\eta \simeq 1/\sqrt{N}$ and this explains the $-1/3$ scaling
 clearly seen in Fig.~\ref{fig:fily}(b).

The same argument, however, does not apply to the oscillator
 with energy near $e_{\rm s}$, because here the instability is
 rather due to the separatrix.
This results in a peak in $\lambda_0(h)$ at the separatrix energy,
 which does not decay as $N^{-1/3}$.
We investigate the scaling behavior of its width, by plotting in
Fig.~\ref{fig:fily}(c) the LE normalized
 by its maximum value $\lambda_{\rm M}$ (for any given $N$)
 with a rescaled axis $(h - e_{\rm s})N^{1/2}$.
The nice overlap of the curves obtained at different system sizes
 shows that the width decreases as $1/\sqrt{N}$. 
This indicates that the anomalous behavior is exhibited by
$\mathcal{O}(\sqrt{N})$ oscillators located
within the range of the separatrix energy.
These are consequences of  
the $\mathcal{O}(1/\sqrt{N})$ fluctuations of the magnetization.

\begin{figure}[t]
\includegraphics[draft=false,clip=true,width=\hsize]{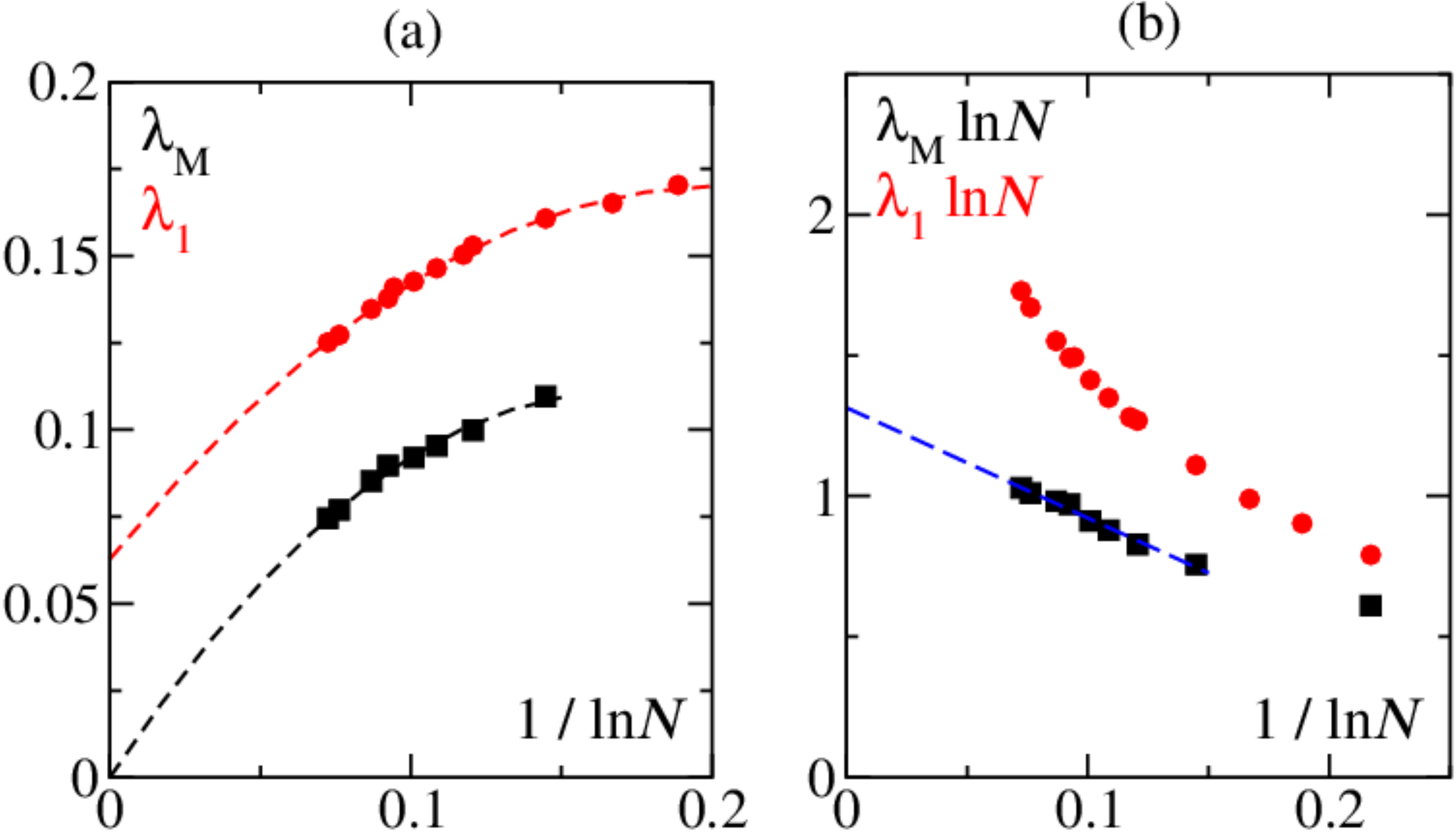}
\caption{(color online). Comparison between the maximum LE $\lambda_{\rm M}$ of a single forced
oscillator (black squares) at separatrix energy $e_{\rm s}$ and the first LE $\lambda_1$ of the
full HMF model for $U=0.5$ (red circles).  The dashed lines highlight the quadratic behavior
$\lambda_1, \lambda_{\rm M} \sim b_0 + b_1/\ln N + b_2/(\ln N)^2$ with $b_0 = 0$ for $\lambda_{\rm M}$.  (a) The LEs versus
$1/\ln N$.
(b) The LEs are multiplied by $\ln N$. The absence of the constant term $b_0$ is confirmed by the linearly arranged symbols for $\lambda_{\rm M}$.}
\label{fig:fily_max}
\end{figure}

Finally, in Fig.~\ref{fig:fily_max} we plotted $\lambda_{\rm M}$ for different
values of $N$ (see black squares). 
The data shows that $\lambda_{\rm M}$ scales as $1/\ln N$ for large $N$. 
This behavior can be understood by introducing a suitable
symbolic dynamics. The main source of uncertainty (and thus of entropy) is
associated to the binary ``choice'' made by the oscillator on reaching
the top of the potential, between the option
to return to the same side or to pass it.
%
Accordingly, we expect
 the metric entropy to be $K \approx \ln 2/t_s$, where $t_s$ is
the return time to the saddle \cite{Note5}. The return time
can be estimated as the time needed to amplify a distance from the
saddle, in the order of the noise amplitude $1/\sqrt{N}$, to a value of
order 1. 
Since the separation rate from the saddle is finite in the thermodynamic limit,
 the condition reads $\exp(t_s)/\sqrt{N} \approx 1$. 
Accordingly, $t_s \approx \ln N$ and therefore $K \approx 1/\ln N$. 
Since the metric entropy is generically estimated
 by the sum of the positive LEs,
 which is simply equal to the sole positive LE in our case,
 we can finally conclude that $\lambda_{\rm M} \approx 1/\ln N$, too. This
prediction is confirmed in Fig.~\ref{fig:fily_max}
 with a quadratic correction for finite $N$, i.e.,
 $\lambda_{\rm M} \approx b_1/\ln N + b_2/(\ln N)^2 + \dots$.
Note that, in contrast to the first LE $\lambda_1$ of the full system
 (red circles in Fig.~\ref{fig:fily_max}),
 $\lambda_{\rm M}$ vanishes in the infinite-size limit as it should be
 (black squares).

By further comparing the single-oscillator LE $\lambda_{\rm M}$
 with the first LE $\lambda_1$ of the full system
 (see red circles in Fig.~\ref{fig:fily_max}), we see that such a
single oscillator contribution shows a size-dependence similar
to that of the full-system LE. However,
for increasing $N$ there is no evidence that the gap is going to close.
Indeed, Fig.~\ref{fig:fily_max}(b) shows that $\lambda_1 \ln N$
diverges for $N \to \infty$, suggesting that 
in the thermodynamic limit the relevant, non vanishing contribution
to the first LE $\lambda_1$ arises from the coupling terms. 
It is therefore necessary to
consider more carefully the whole evolution equation in tangent space
and the role played in this context by the fluctuations of the finite-time LE.\\
A simplified argument can be put forward to infer the asymptotic behavior
 of the fluctuations of the finite-time LE for a forced oscillator near the separatrix energy.
Whenever the oscillator passes near the saddle,
 its growth rate is always of order 1, positive or negative,
 irrespective of the system size.
This implies that the amplitude of the finite-time LE fluctuations
 near the separatrix energy should remain
 positive in the thermodynamic limit,
 possibly with logarithmic finite-size corrections.
In the next section, we will analyze these fluctuations more closely
in order to build a simplified model for the fully coupled tangent
space dynamics which will allow us to establish a finite lower bound for the largest LE.

\section{A simplified model for the coupling pressure}
\label{SecToy}

In the context of globally-coupled dissipative systems, we recently showed
that global coupling may induce an increase of the first LE
 with respect to the single-unit exponent \cite{short}. Here, we
refine such argument for the HMF context.

The single oscillator approximation discussed in the previous section consists
in disregarding the contribution of the coupling term
 appearing in the second line of Eq.~(\ref{eq:tang1}). 
In fact, as we explain below,
 this is what happens to the tangent-space evolution of the full system
 for most of the time, because of the localization of the Lyapunov vector
 (see Fig.~\ref{fig4}) and
the $1/N$ normalization in front of the coupling term. 
The Lyapunov vector components then evolve independently. 
In particular, the logarithms of the
amplitudes ($x_i=\ln \sqrt{A_i}$) can be regarded as Brownian particles with a drift
velocity given by the single-particle LE and an effective diffusion constant that measures
the fluctuations of the LE itself. The analysis carried out in the previous
section suggests that the oscillators should be classified into two groups: 
(i) a
small fraction which lies close to the separatrix and is characterized
by a LE of order $1/\ln N$ and non zero fluctuations of
  the finite time LE; 
(ii) the vast majority, characterized by a LE of order $N^{-1/3}$
and vanishing fluctuations. 
The $\mathcal{O}(1/\sqrt{N})$ fluctuations of the magnetization
 then suggest that the population ratio
between the particles close to the separatrix and the remaining
population vanishes in the thermodynamic limit,
possibly as $1/\sqrt{N}$.
Moreover, energy diffusion induces (slow)
exchanges between the two families.\\
We now discuss how the coupling modifies the single oscillator evolution.
The localization of the Lyapunov vector indicates that
 the coupling term (the sum in the
r.h.s. of Eq.~(\ref{eq:tang1})) is of the order of $\delta \theta_m/N$, where
$m$ labels the oscillator where the vector is localized. Therefore, because of
the $1/N$ factor, the $m$th oscillator component only weakly affects the oscillator at stake, $\delta \theta_i$, and so does the coupling term. 
However, the opposite is true when
$|\delta \theta_i| \ll \ |\delta \theta_m|/N$ (notice that this is possible,
since the various components evolve independently of each other, and their
logarithms diffuse away). 
In this latter case, the evolution is dominated by the
coupling and the net result is that such extremely small components become
of the order of the coupling term. In terms of the logarithmic coordinates
$x_i$, the effect of the coupling can be schematized by a barrier
sitting at $x_{\rm min} = x_{\rm max} - \ln N$ 
(where $x_{\rm max}$ labels the rightmost
particle, that is the largest vector coponent) which prevents any
interparticle distance from being larger than $\ln N$. 

\begin{figure}
\includegraphics[draft=false,clip=true,width=\hsize]{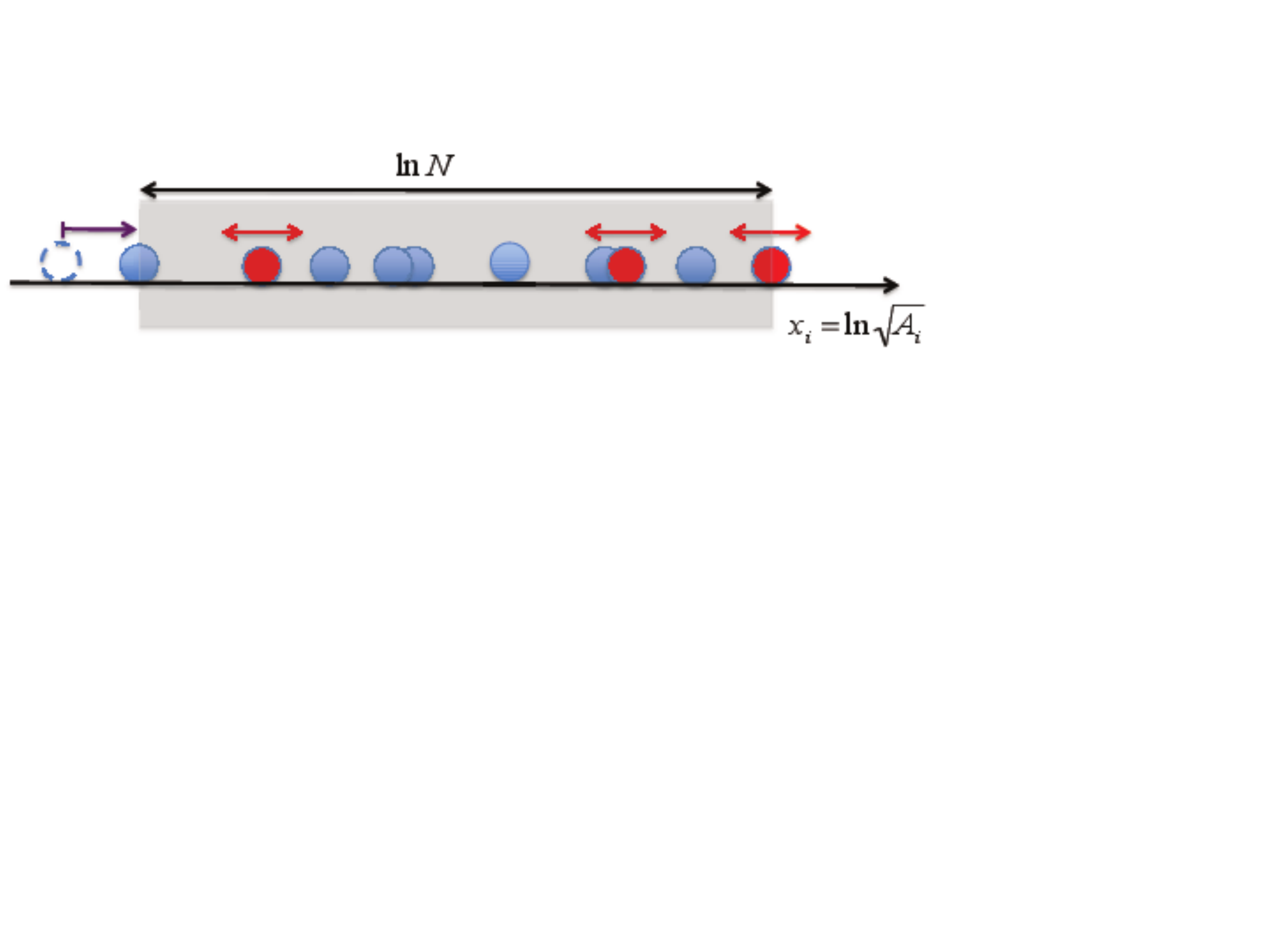}
\caption{(color online). Schematic representation of the toy model for the coupling pressure (see text). Point-like particles sit at the logarithmic coordinates $x_i=\ln\sqrt{A_i}$ and are either diffusive (red circles) or non diffusive (blue). Diffusive particles turn non diffusive with a rate $\alpha_1$ and non diffusive ones start to diffuse with a rate $\alpha_2 \sim \alpha_1/\sqrt{N}$, as the corresponding oscillators approach and leave the separatrix, respectively. The coupling is zero as long as particles are no farther than $\ln N$ from the rightmost particle, but otherwise acts as a barrier, preventing particles from being farther than $\ln N$ from the rightmost one. The net effect is a drift to the right.}
\label{fig-ToyModel}
\end{figure}

Altogether, we propose a simplified model of two populations of ``particles'',
 as sketched in Fig.~\ref{fig-ToyModel}.
The particles in the first group (red in Fig.~\ref{fig-ToyModel})
 show the biased Brownian motion,
 while those in the other group (blue) stay quiescent.
Each particle evolves independently of the others until it lies 
at a distance larger than $\ln N$
to the left of the rightmost particle (where the Lyapunov vector is localized),
in which case it is instantaneously pushed forward by the coupling to restore
the maximal allowed distance
 (drawn by the gray box of size $\ln N$ in Fig.~\ref{fig-ToyModel}). 
A precise formalization of the model
requires the following additional ingredients: drift velocities and diffusion
coefficients of the two populations and the mutual transition rates.
For the drift, we assume that both populations are characterized by a zero
velocity (zero LE). Since our goal is to explain the origin of
a strictly positive LE in the thermodynamic limit, we believe that neither
a $1/N^{1/3}$ nor a $1/\ln N$ LE can eventually provide a leading contribution
and thereby set both to be zero. As for the diffusion
coefficient of the Brownian particles,
we assume a finite value $D_{\rm s}$ for the first population (that corresponding to 
the finite-time LE fluctuations of the oscillators
 in the vicinity of the separatrix). In contrast, we
assume a zero diffusion coefficient for the second population, as
it is negligible for large sizes. As for the transition
rates $\alpha_1$ and $\alpha_2$ from the diffusing to the still population and
vice versa, respectively, on the basis of the numerical observation
 and the theoretical argument on the ratio of the two populations,
 we assume that the ratio $\alpha_2 / \alpha_1$ vanishes in
the thermodynamic limit as $1/\sqrt{N}$. 
Thus, we are left with three independent parameters: the diffusion
coefficient $D_{\rm s}$, the transition rate from the diffusing to the
still population $\alpha_1$, plus a small parameter
 $\alpha_2$ of the order of $\alpha_1/\sqrt{N}$.

It is convenient to introduce the probability density $P_j(x,t)$ for a particle
of the $j$th population ($j=1$ and 2 referring to the diffusing and still
populations, respectively) to be at position $x$. 
If 
both populations move with a positive velocity $v$, 
 a positive LE spontaneously emerges
 in the system of interacting particles. It is convenient
to study the problem in a frame moving with a velocity $v$, since then
one has to look for a stationary solution. In this frame,
the evolution equation reads,
\bey
\frac{\partial P_1}{\partial t} &=& v \frac{\partial P_1}{\partial x} +
\frac{D_{\rm s}}{2}  \frac{\partial^2 P_1}{\partial x^2} - \alpha_1 P_1 +
\alpha_2 P_2, \nonumber \\
\frac{\partial P_2}{\partial t} &=& v \frac{\partial P_2}{\partial x} 
+ \alpha_1 P_1 - \alpha_2 P_2,
\label{Toy2}
\eey
where $x\in [0,\infty]$, with a reflecting boundary at $x=0$. 
Equation (\ref{Toy2})
describes an ensemble of stochastic particles that move along a
tilted plane, which corresponds to the velocity $v$ of the comoving frame,
 and have two possible internal states, one
characterized by a finite diffusion $D_{\rm s}$
 and the second one by a zero diffusion,
so that the particles in the second state
 simply move toward the reflecting barrier at $x=0$. 
The diffusive dynamics competes with the time scales set by the
transition rates between the two populations. In particular, 
if $D_{\rm s}$ is finite and 
$\alpha_1$ and $\alpha_2$ are sufficiently small,
particles in the non diffusing populations will
tend to accumulate in $x=0$. Therefore, we look for a general stationary
solution of Eq.~(\ref{Toy2}) of the form
\bey
P_1(x)&=&c_1 e^{-\gamma x}, \nonumber\\
P_2(x)&=&c_0 \delta(x) + c_2 e^{-\gamma x},
\label{ansatz}
\eey
with some $\gamma>0$ and Dirac's delta $\delta(x)$.
The two probability densities must obey particle conservation
\be
\int_0^{\infty} \left ( P_1(x) + P_2(x)\right)\, dx = 1,
\label{b1}
\ee
and the population equilibrium condition
\be
\alpha_1 \int_0^{\infty} P_1(x)\, dx = \alpha_2 \int_0^{\infty} P_2(x)\, dx.
\label{b2}
\ee
Substituting the Ansatz (\ref{ansatz}) into Eq.~(\ref{Toy2}), we obtain
the stationary conditions in the bulk ($x \neq 0$),
\bey
0 &=& -v c_1 \gamma + \frac{D_{\rm s}}{2} \gamma^2 c_1 -\alpha_1 c_1 +
\alpha_2 c_2, \nonumber \\
0 &=& -v c_2 \gamma +\alpha_1 c_1 -\alpha_2 c_2,
\eey
which yields (for $v \neq 0$)
\be
c_2 = c_1 \left(\frac{D_{\rm s} \gamma}{2 v} -1 \right),
\label{b0}
\ee
and
\be
2 \gamma v^2 + \left[ 2 (\alpha_1 + \alpha_2) - D_{\rm s} \gamma^2 \right] v
- D_{\rm s} \alpha_2 \gamma = 0.
\label{b3}
\ee
This can be solved for $v$, choosing the physically meaningful
 positive solution.
By recalling that
$\alpha_2 \sim \alpha_1/\sqrt{N}$, we can expand in terms of $\alpha_2$,
\be
v = \left\{
\begin{array}{lr}
\frac{D_{\rm s}}{2} \gamma - \frac{\alpha_1}{\gamma} + 
 {\mathcal O}\left(\alpha_2\right) &
\;\;\mbox{if} \quad (\alpha_1+\alpha_2) <  D_{\rm s} \gamma^2 /2,\\
\frac{D_{\rm s} \gamma}{2\alpha_1 - D_{\rm s} \gamma^2} \alpha_2 - 
  {\mathcal O}\left(\alpha_2^2\right)
& \;\;\mbox{if}\quad (\alpha_1+\alpha_2) >  D_{\rm s} \gamma^2 /2.
\end{array}\right.
\label{eq:v}
\ee
Note that the velocity (i.e., the LE) is
strictly positive when diffusion dominates over the interstate
transitions.

From Eqs.~(\ref{ansatz})-(\ref{b2}) it also follows 
\be
c_1 = \gamma \left(\frac{\alpha_2
  /\alpha_1}{1+\alpha_2/\alpha_1}\right) \sim \gamma\frac{\alpha_2}{\alpha_1},
\label{b4}
\ee
and together with Eqs.~(\ref{b0}) and (\ref{eq:v}),
 we find that the coefficient of Dirac's delta has a finite amplitude,
\be
\frac{c_0}{2} = \left\{\begin{array}{lr}
1 - {\mathcal O}\left(\alpha_2\right)&\;\;\mbox{if}\quad
(\alpha_1+\alpha_2) < D_{\rm s} \gamma^2 /2,\\
\frac{D_{\rm s} \gamma^2}{2 \alpha_1} -{\mathcal
  O}\left(\alpha_2\right) & \;\;\mbox{if}\quad (\alpha_1+\alpha_2)
>  D_{\rm s} \gamma^2 /2.
\end{array}\right.
\label{eq:c0}
\ee

We can now determine $\gamma$ self-consistently. 
Given that we have an ensemble of $N$ particles
 whose rightmost position is $x_{\rm max}$ ($= \ln N$),
the integrated probability in the excess region,
$\int_{x_{max}}^\infty (P_1(x) + P_2(x))\,dx$, should be in the order of $1/N$.
From Eqs.~(\ref{ansatz}) and (\ref{b0}) we have
\be
e^{-\gamma x_{max}} = \frac{2 v}{D_{\rm s} c_1}\frac{d_0}{N}  \, ,
\label{b5}
\ee
where $d_0$ is a constant of ${\mathcal O}(1)$.
By substituting Eqs.~(\ref{eq:v}) and (\ref{b4})
into Eq.~(\ref{b5}) and using $\alpha_2/\alpha_1 \sim 1/\sqrt{N}$,
 we obtain, for small transition rates ($\alpha_1< D_{\rm s} \gamma^2/2 $),
\be
e^{-\gamma x_{max}} = \tilde{d}_0 \left( 1 - \frac{2 \alpha_1}{D_{\rm s}
    \gamma^2}\right) \frac{1}{\sqrt{N}} + {\mathcal O}\left(\alpha_2^2\right),
\label{b6}
\ee
where $\tilde{d}_0$ is another $\mathcal{O}(1)$ constant.

\begin{figure}[t]
\includegraphics[width=0.9\hsize,clip]{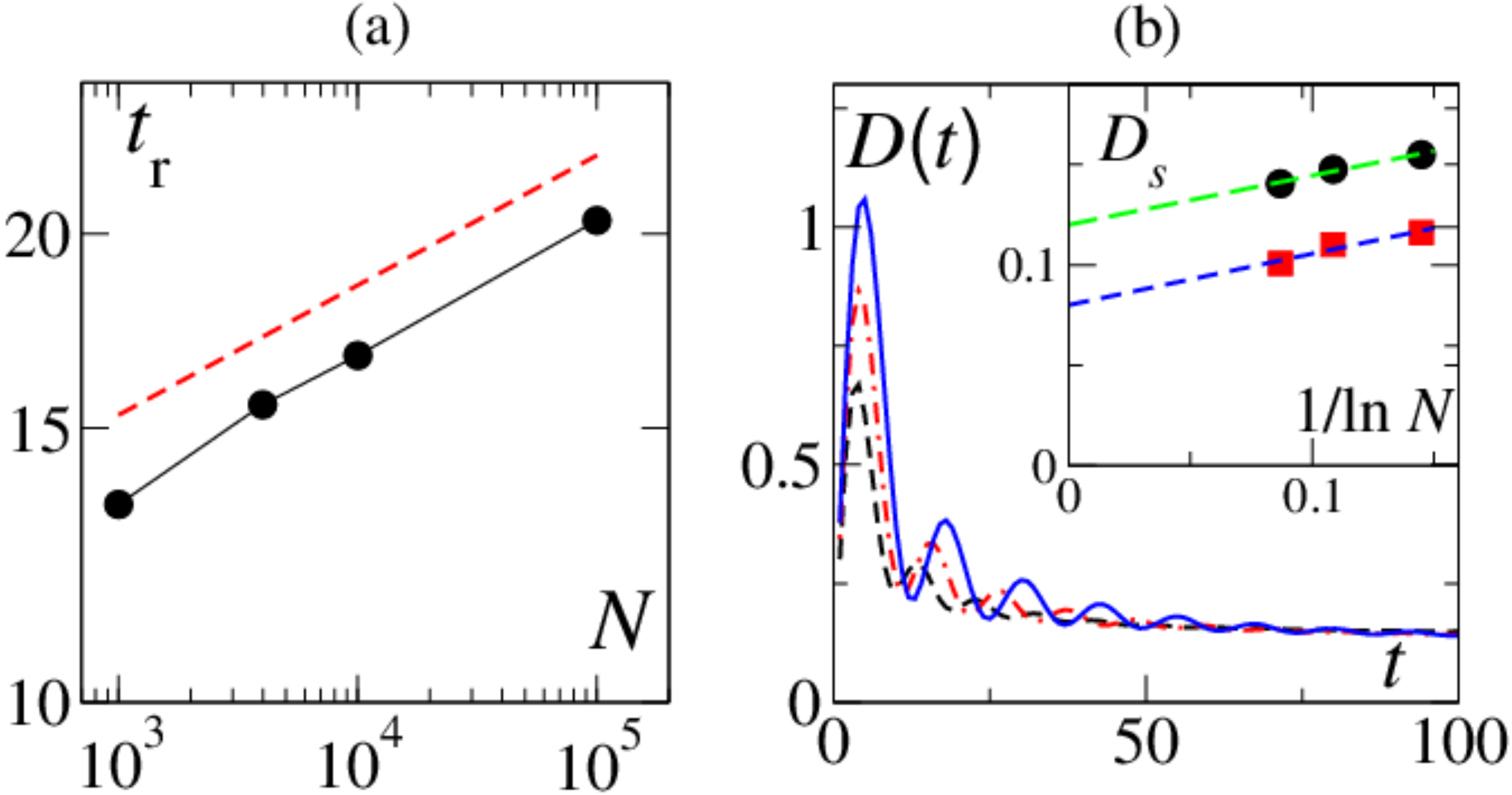}
\caption{(color online). (a) Average single-particle residence time $t_{\rm r}$ near
the separatrix vs $N$ for $U=0.5$. The residence time is estimated by
measuring the crossing time needed to a particle to pass
from an energy $e_{\rm L}$ to an energy $e_{\rm R}$ or vice versa,
 where $e_{\rm L} < e_{\rm R}$ are the two energies corresponding to the half
maximum of the curve $\lambda_0(h)$ reported in Fig.~\ref{fig:fily}.
The average is taken over $5 \times 10^5$ to $2 \times 10^6$
events for each size $N$. The dashed red line indicates $N^{1/12}$.
(b) Single-particle finite-time diffusion $D(t)$
vs. time $t$ for $U=0.5$. 
Dashed (black), dot-dashed (red) and solid (blue) lines correspond 
corresponds to $N=10^3$, $N=10^4$, and $N=10^5$, respectively.
Inset: Effective diffusion coefficient $D_{\rm s}$ (see text) as a
function of $1/\ln N$. Black circles and red squares correspond
to $U=0.5$ and $U=0.7$, respectively. The dashed lines
mark the linear extrapolation to the asymptotic value.}
\label{fig:fpt}
\end{figure}

In the HMF, the transition rate $\alpha_1$ is the inverse of the
residence time $t_r$ of an oscillator
 near the separatrix energy, whose width has been shown to
 scale as $1/\sqrt{N}$. We now compute the scaling
behavior of this residence time, analyzing more closely the dynamics
near the energy maximum $\theta - \phi_N = \pi$. Consider a particle
with energy $2 M_N$ and
 phase space coordinates $p=0$ and $\theta=\phi_N+\pi$. 
By expanding Eq.~(\ref{H1}) around
the potential energy maximum,
 we obtain
\be
h\simeq\frac{p^2}{2}+2 M_N-\frac{M_N}{2} \left(\theta-\phi_N - \pi\right)^2
\label{H1:exp}
\ee
and the following equations for the single particle dynamics
\begin{align}
&\dot\theta = p,  \nonumber \\
&\dot p \simeq M_N \left (\phi_N - \theta + \pi\right).
\label{dynamics:exp}
\end{align}
We already know from Sec. \ref{subsec:magn} that the global phase $\phi_N$ exhibits a drift
 $\Delta \phi = \phi_N(t)-\phi_N(0) \simeq \omega t$ on timescales smaller than
$\tau_{\rm diff} \sim N$.
By integration, we find that 
$\Delta \theta = \theta(t)-\theta(0) \sim \omega t^3$ which dominates
the dynamics of the single particle energy for large times, as
\be
\Delta h \sim \Delta \theta^2 \sim \omega^2 t^6.
\ee
By finally recalling that $\omega \sim 1/\sqrt{N}$, we can determine the scaling
of the time needed for $\Delta h$ to grow up to order $1/\sqrt{N}$, that is
\be
\frac{1}{\alpha_1} = t_r \sim N^{1/12}\,,
\ee
which is in agreement with numerical results reported in
Fig.~\ref{fig:fpt}(a).

As a result, $\alpha_1$ goes to zero algebraically (albeit with a small
exponent) in the thermodynamic limit
and this guarantees that the inequality $\alpha_1< D_{\rm s} \gamma^2/2$ is
always satisfied asymptotically.
By further imposing that $x_{\rm max}$ equals to the box width
$\ln N$, we
have from Eq.~(\ref{b6})
\be
\gamma = \frac{1}{2} + {\mathcal O}\left(\frac{1}{\ln N}\right).
\label{eq:gamma}
\ee

By now substituting Eq.~(\ref{eq:gamma}) into Eq.~(\ref{eq:v}), we finally
obtain 
\be
v = \frac{D_{\rm s}}{4} + {\mathcal O}\left(\frac{1}{\ln N}\right) \, .
\label{EqMainResult}
\ee
In other words,
 because of the divergence of the residence time near the separatrix,
 which induces sustained fluctuations of the finite-time LE,
 the asymptotic value of the velocity, or the (time-averaged) LE is finite. 
From the quantitative point of view, it is
important to notice that our estimation for the asymptotic LE
\begin{equation}
\lambda_\infty = \frac{D_{\rm s}}{4}
\label{EqMainResult2}
\end{equation}
should be interpreted as a lower bound for the actual LE. 
In fact, 
it has been obtained by assuming the vanishing mean velocity
 for both of the populations, as well as the vanishing fluctuations
 for the still one.
Accordingly, there are reasons to expect that the contribution to the LE 
from the coupling term may be larger than $D_{\rm s}/4$.

In order to compare our estimates
 with the asymptotic values $\lambda_{\infty}=0.056(6)$ ($U=0.5$)
and $\lambda_{\infty}=0.046(3)$ ($U=0.7$) (see Sec.~\ref{Lyapunov}),
 we need to estimate the effective diffusion coefficient $D_{\rm s}$
 for a single forced oscillator lying at the separatrix energy $h=e_{\rm s}$.

From time series $\sim 10^6$ time units long, we determine the
mean-square displacement of the integrated Lyapunov exponent and thereby, 
after dividing by the elapsed time $t$, the finite-time diffusion coefficient $D(t)$
that is shown in Fig. ~\ref{fig:fpt}(b) for $U=0.5$ and three different 
numbers $N$ of oscillators (which contribute to the magnetization). 
There we notice that, upon increasing $N$, $D(t)$ exhibits increasing
oscillations of increasing period. This is a manifestation of the presence
of long stretches of positive (negative) local exponents when the 
oscillator is located close to the saddle. In fact, the period is 
proportional to $t_s\approx \ln N$. On the other hand, the effective 
diffusion coefficient $D_s$ should be estimated on a time scale of the 
order of the residence time $t_r$ close to the separatrix and, since
we have just seen that $t_r \approx N^{1/12}$, it turns out that in
the thermodynamic limit $t_r>>t_s$.
In the lower-bound spirit of our estimates, we choose to identify $D_{\rm s}$ with 
the minimum of the finite-time diffusion $D(t)$ in the time interval $t\in [0, 5 t_r]$. 
The results are shown in the inset of Fig.~\ref{fig:fpt}(b) for both $U=0.5$ and $U=0.7$.
By varying the number $N$ of forcing oscillators from $10^3$ to $10^5$ and 
assuming again $1/\ln N$ corrections, we obtain the asymptotic estimates $D_{\rm s}=0.12$ 
for $U=0.5$ and $D_{\rm s}=0.08$ for $U=0.7$. It turns out that there is approximately a
factor two between $D_{\rm s}/4$ and $\lambda_{\infty}$. The main interest of the 
formula (\ref{EqMainResult}) is, however, that, representing a 
lower bound, it shows that the largest LE remains strictly positive
in the infinite-size limit.

\section{Critical behavior of the largest Lyapunov exponent}
\label{sec:critical}

So far we have shown that, in the ordered phase $U < U_{\rm c}$, the coupling
pressure due to oscillators near the separatrix keeps the largest LE
 $\lambda_1$ positive even in the infinite-size limit. This argument does
not hold in the disordered phase $U > U_{\rm c}$, where the magnetization $M$
 (and the separatrix) vanish. Instead, as already mentioned,
 the largest LE decays to zero as $\lambda \sim N^{-1/3}$.

An interesting question arises then quite naturally: what is the behavior of the
largest LE in the vicinity of the critical point $U_{\rm c}$?
The critical behavior of the largest LE $\lambda$ may provide a connection between a
dynamic quantity of the full system and macroscopic thermodynamic properties.

\begin{figure}[t]
\includegraphics[width=\hsize,clip]{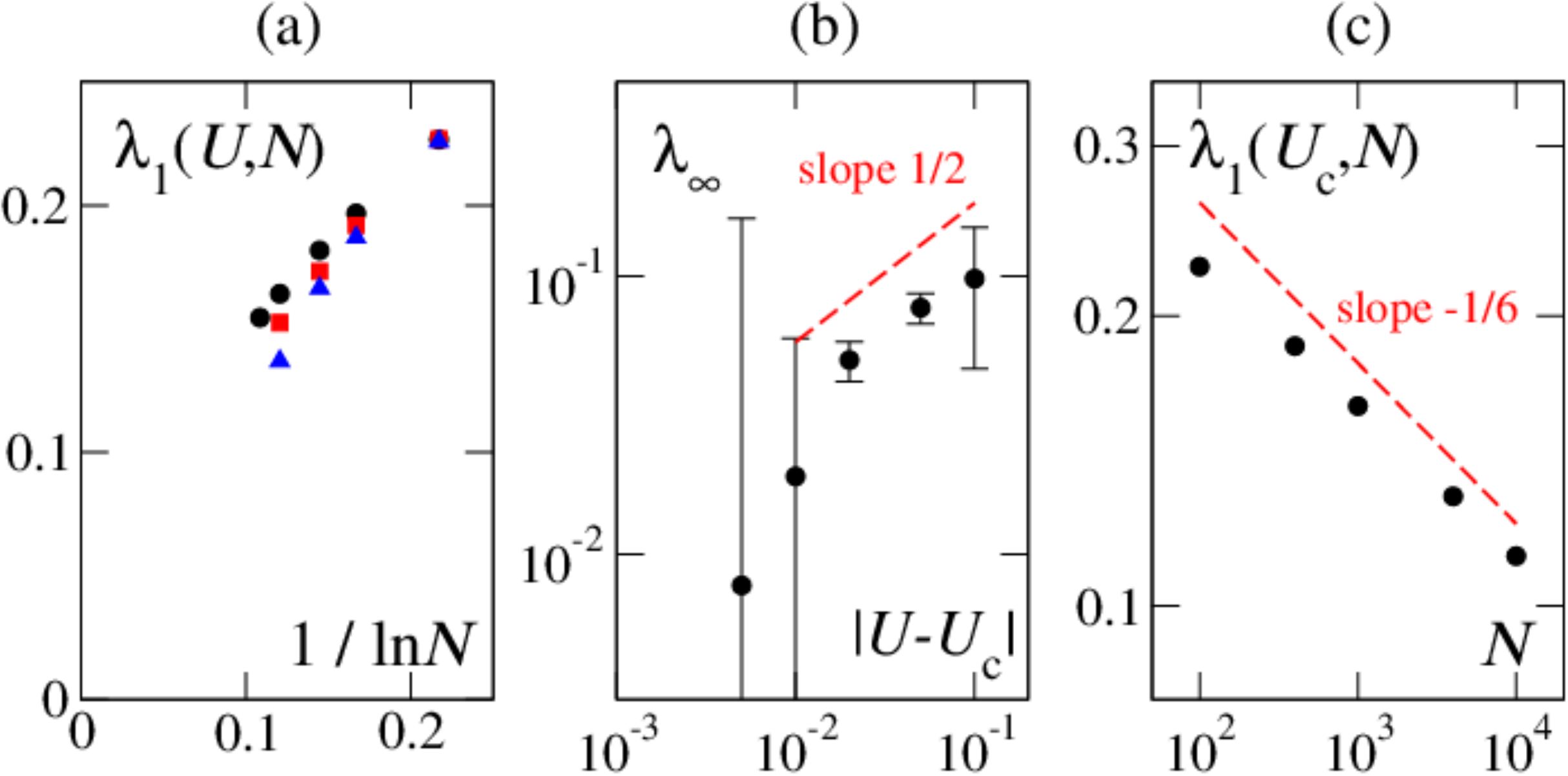}
\caption{(color online). Critical scaling of the largest LE $\lambda_1(U,N)$. 
(a) $\lambda_1$ as a function of $1/\ln N$ for different energies
$U = 0.70, 0.73$, and $0.745$ from top to bottom. (b) Extrapolated values of the asymptotic largest LE $\lambda_\infty$ as a
function of the distance from the criticality $|U-U_{\rm c}|$. (c) Size
dependence of $\lambda_1$ at the critical point $U=U_{\rm c}$.}
\label{fig:crit2}
\end{figure}%

In this section we numerically investigate the critical properties of the
largest LE $\lambda_1(U,N)$ in the HMF model, providing a
theoretical account for the observed finite-size scaling. We have already
seen that, in the ordered phase, $\lambda_1$ decreases logarithmically with
increasing system size, toward a strictly positive asymptotic value
 $\lambda_\infty(U)$. While approaching the critical point, 
however, we find that
the logarithmic decay sets in at larger and larger sizes
[Fig.~\ref{fig:crit2}(a)] and converges to smaller values
 of $\lambda_\infty(U)$.
Although large finite-size effects as well as critical slowing down prevent us from
estimating $\lambda_\infty(U)$ near the critical point, our estimates in
Fig.~\ref{fig:crit2}(b) suggest that the largest LE exhibits
 the same critical scaling as the magnetization
 with respect to the system energy $U-U_{\rm c}$, namely,
\begin{equation}
\lambda_\infty(U) \sim |U - U_{\rm c}|^{1/2},~~~~ \text{for $U < U_{\rm c}$.}
\label{eq:crit6}
\end{equation}
In this context Firpo's Riemannian theory \cite{firpo}
 predicted a different power law
$\lambda_\infty(U) \sim |U-U_{\rm c}|^{1/6}$ for $U < U_{\rm c}$,
 though it was derived under assumptions
 that are not valid near the critical point.

At criticality, the logarithmic dependence of $\lambda$ on $N$ is replaced by
an algebraic decay, $\lambda_1(U_{\rm_c},N) \sim N^{-1/6}$, toward a vanishing
$\lambda_\infty$ [Fig.~\ref{fig:crit2}(c)]. In fact, this behavior can also be
 explained by the random matrix argument
 for the power-law decay $\lambda_1 \sim N^{-1/3}$
 in the disordered phase \cite{rev1,Anteneodo_Vallejos-PRE2001}.
In the latter case, the disorder of the matrices
 is due to the statistical fluctuations of the magnetization,
 and thus its amplitude $\eta$ scales as $1/\sqrt{N}$
 and the largest LE $\lambda_1 \sim \eta^{-2/3} \sim N^{-1/3}$.
By contrast, at the critical point, the disorder is due to
 the critical decay of the magnetization
 $M(U_{\rm c}, N) \sim N^{-\beta/\nu}$ (see Sec.~\ref{subsec:magn}),
 which yields
\begin{equation}
 \lambda(U_{\rm_c},N) \sim N^{-2\beta/3\nu}.  \label{eq:crit7}
\end{equation}
By recalling $\beta =1/2$ and $\nu=2$, this indicates
$\lambda(U_{\rm_c},N) \sim N^{-1/6}$, in agreement with the numerical
observation in Fig.~\ref{fig:crit2}(c).

\section{The Full Lyapunov Spectrum}
\label{sec:full}
In this section we study the Lyapunov spectrum of the HMF model in the
ordered phase. This analysis provides a more detailed characterization of
the instability. In particular it helps to assess the (non)extensivity of the
chaotic dynamics. Given the difficulty of extending the theoretical arguments
 in Sec.~\ref{SecToy} beyond the largest LE,
 we restrict our studies to a careful
numerical analysis. Given the symmetry of the Lyapunov spectrum in
Hamiltonian systems, it is sufficient to compute the first half.

In Ref.~\cite{tanos} it has been argued that, for $U=0.1$, the full spectrum
vanishes roughly as $N^{-1/3}$. However, the intermittent behavior observed at
low energies [see, e.g., \figref{fig2}(b-d)] indicates that the burst state should
eventually (for $N$ large enough) dominate and thus
 the $N^{-1/3}$ law eventually breaks down.
Given this difficulty of dealing with low energy values,
 we focus here on a larger energy value, namely $U=0.7$.

\begin{figure*}[t]
\begin{center}
\includegraphics[width=\hsize,clip]{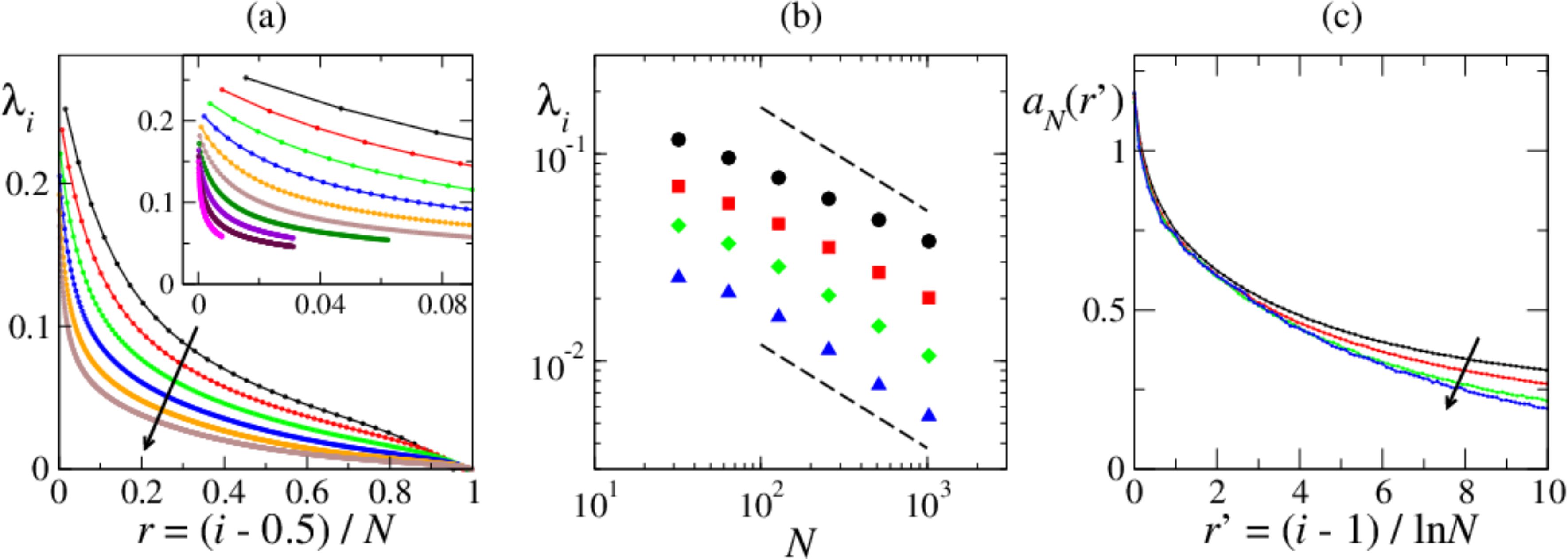}
\caption{Full Lyapunov spectrum at $U = 0.7$. (a) Lyapunov spectrum
$\lambda_i$ as a function of the rescaled index $r \equiv (i-0.5)/N$ for
system sizes $N=32, 64, \cdots, 1024$ from top right to bottom left, as
indicated by the arrow. Inset: close-up of the first part, with sizes $N=2048,
4096, \cdots, 16384$ added. (b) $\lambda_i$ vs $N$ at fixed rescaled indices
$r = 0.2, 0.4, 0.6$, and $0.8$ (from top to bottom). The dashed lines indicate
$\lambda_i \sim 1/\sqrt{N}$. (c) $a^{(N)}(r')$, as defined in Eq.
(\ref{EqFullSpecSub2}), plotted as a function of the logarithmically rescaled index $r'=(i-1) / \ln N$ for $N=1024, 2048, \cdots, 16384$. They show reasonable behavior toward the convergence, implying the logarithmic size-dependence (\ref{EqFullSpecSub}) for these subextensive LEs (see text).}
\label{fig-FullSpec}
\end{center}
\end{figure*}

Figure \ref{fig-FullSpec}(a) shows the Lyapunov spectra $\lambda_i$
as functions of the rescaled index $r \equiv (i-0.5)/N$
for different system sizes $N$. 
This suggests that the spectrum is composed of two parts: 
the bulk of the spectrum which decays toward zero for increasing
$N$, and the initial part pinned close to the largest LE,
 which is clearly visible in the inset of Fig.~\ref{fig-FullSpec}(a). 

This scenario is actually coherent with the coexistence of extensive and
subextensive chaos, recently discovered in generic globally-coupled
dissipative systems \cite{short}. In such systems, the Lyapunov spectrum is
found to be asymptotically flat (a specific realization of extensivity)
 but sandwiched
between two vanishing fractions of exponents located at both ends of the
spectrum, where different asymptotic values appear. Finite-size analysis
 then revealed that the bulk of the spectrum scales as
\begin{equation}
 \lambda_i \simeq \lambda_0 + \frac{c(r)}{\sqrt{N}}
\label{EqFullSpecBulk}
\end{equation}
where the asymptotic value $\lambda_0$ corresponds to the LE
 of a single dynamic unit forced by the mean field \cite{short}.

In \figref{fig-FullSpec}(b), one can appreciate that the spectrum of the HMF
decays as predicted by Eq.~(\ref{EqFullSpecBulk}) with $\lambda_0 = 0$, since
a single oscillator has only two variables and thus cannot be
chaotic. Notice that the existence of these zero-Lyapunov bulk components is
consistent also with the theoretical prediction of Ref.~\cite{kurchan}, which
did not exclude the presence of a vanishing (subextensive) fraction of
different exponents. As for the power-law decay of the bulk exponents, the data
for small $r$-values in \figref{fig-FullSpec}(b) seem to decrease more
slowly than $1/\sqrt{N}$, but this is presumably due to strong finite-size
corrections induced by the bending near the beginning of the spectrum.

Concerning the subextensive LEs, in dissipative systems it was found
that there are $\mathcal{O}(\ln N)$ exponents whose values vary as
$\lambda^{(i)} \simeq \lambda_\infty + a(r')/\ln N + \mathcal{O}(\ln^2 N)$
 with $\lambda_\infty \neq \lambda_0$ independent of $r'$, when one fixes a
logarithmically rescaled index $r' \equiv (i-1)/(i_0 + \ln N)$ with a constant
$i_0$ \cite{short}. 
In the HMF, we have shown both numerical (\figref{fig2}) and
theoretical (Sec.~\ref{SecToy}) evidence of this logarithmic dependence 
for the largest LE, 
$\lambda_1 \simeq \lambda_\infty + a(0) / \ln N + b(0)/\ln^2 N$,
with $\lambda_\infty = 0.046(3)$ for $U=0.7$.
Now, we assume that the same size-dependence holds for subsequent LEs
 like in dissipative systems, with varying coefficients except
 for the constant term:
\begin{equation}
 \lambda_i \simeq \lambda_\infty + a(r') / \ln N + b(r')/\ln^2 N.
 \label{EqFullSpecSub}
\end{equation}
To examine the validity of this expression, we take
 the Lyapunov spectra $\lambda^{(N)}(r') \equiv \lambda_i^{(N)}$
 at system size $N$ and compute
\begin{equation}
 a^{(N)}(r') \equiv \frac{\Delta\lambda^{(2N)}(r')\ln^2 2N - \Delta\lambda^{(N)}(r')\ln^2 N}{\ln 2},
 \label{EqFullSpecSub2}
\end{equation}
 with $\Delta\lambda^{(N)}(r') \equiv \lambda^{(N)}(r') - \lambda_\infty$.
If Eq.~(\ref{EqFullSpecSub}) holds,
 the definition in Eq.~(\ref{EqFullSpecSub2}) gives
 $a^{(N)}(r') \simeq a(r')$ and the size-dependence vanishes.
Figure \ref{fig-FullSpec}(c) tests this idea
 and indeed verifies that $a^{(N)}(r')$ approaches an asymptotic curve
 for large sizes $N$
 with the logarithmically rescaled index $r' = (i-1)/ \ln N$
 (here $i_0$ is set to be zero).
Therefore, the logarithmic size-dependence (\ref{EqFullSpecSub})
 holds for these subextensive exponents, similarly to dissipative systems.
Although we need to study larger
systems to obtain a firmer numerical support, our results on the full spectrum
of the HMF model are consistent with the coexistence of extensive and
subextensive exponents, previously found for dissipative systems.

\section{The generalized HMF model}
\label{sec:generalized}

In order to study the generality of our results, we finally turn our attention
to a two-dimensional variant of the HMF model, introduced by Antoni
and Torcini \cite{AntoniTorcini} and later generalized \cite{art,rev1}
 to the present
form. It is defined by the Hamiltonian
\begin{widetext}
\begin{equation}
 H = \sum_{i=1}^N \frac{p_{x,i}^2 + p_{y,i}^2}{2} +
\frac{1}{2N} \sum_{i,j=1}^N \left\{ \left[1-\cos(x_i-x_j)\right] + \left[1-\cos(y_i-y_j)\right] +
A\left[1-\cos(x_i-x_j)\cos(y_i-y_j)\right] \right\},
\label{hmf2}
\end{equation}
with two-dimensional coordinates $(x_i,y_i)$, their conjugate momenta
$(p_{x,i}, p_{y,i})$, and a coupling constant $A$.
The equations of motion can be written as,
\begin{align}
 &\dot{x}_i = p_{x,i}, \notag \\
 &\dot{y}_i = p_{y,i}, \notag \\
 &\dot{p}_{x,i} = -M_x \sin(x_i - \phi_x) - \frac{A}{2} \left[ P_+ \sin(x_i+y_i - \psi_+)
  + P_- \sin(x_i-y_i - \psi_-) \right],  \notag \\
 &\dot{p}_{y,i} = -M_y \sin(y_i - \phi_y) - \frac{A}{2} \left[ P_+ \sin(x_i+y_i - \psi_+)
  - P_- \sin(x_i-y_i - \psi_-) \right],  \label{EqMotionHMF2d}
\end{align}
with four mean field terms defined as
\begin{equation}
M_z \e^{\ri\phi_z} \equiv \frac{1}{N}\sum_{i=1}^N \e^{\ri z_i}~~z=\{x,y\},
~~~~~~~~P_\pm \e^{\ri\psi_\pm} \equiv \frac{1}{N}\sum_{i=1}^N \e^{\ri (x_i \pm y_i)}\,,
\label{EqOrderParamHMF2d}
\end{equation}
As a matter of fact, because of the symmetries of the model, on average
$M_x \sim M_y \sim M$  and $P_+ \sim P_{-} \sim P$ and the model can be
described in terms of two order parameters only.
The single oscillator energy can then be written as 
\begin{equation}
h_i = \frac{p_{x,i}^2 + p_{y,i}^2}{2}
+2 +A 
- M[\cos(x_i - \phi_x) + \cos(y_i - \phi_y)]
- \frac{AP}{2} [\cos(x_i+y_i - \psi_+) + \cos(x_i-y_i - \psi_-)]
\end{equation}
(note that in this Section we have not fixed the ground state energy at zero).\\
\end{widetext}

\begin{figure}[t]
\begin{center}
\includegraphics[width=\hsize,clip]{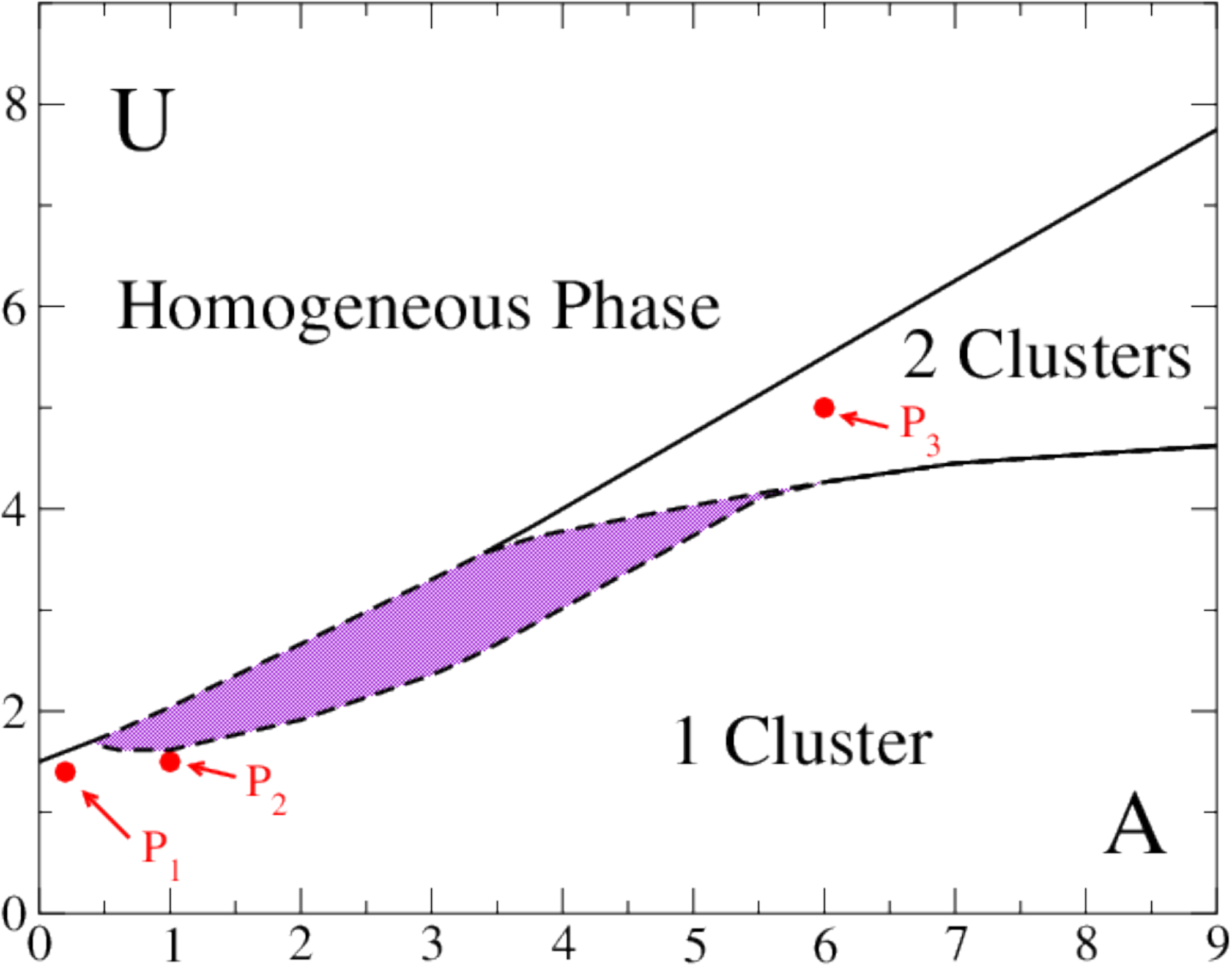}
\caption{(color online). Phase diagram of the generalized HMF model. 
The shaded area denotes the region where microcanonical and canonical ensembles differ 
from one another (i.e. the coexistence region of two different thermodynamic phases) 
occurring in correspondence of discontinuous canonical transitions. Solid
and dashed lines correspond to continuous and discontinuous transitions,
 respectively,
within the canonical ensemble. The red dots indicate the three parameter values we have studied.}
\label{FigPhaseDiagram2d}
\end{center}
\end{figure}

This generalized HMF model is known for its rich and generic behavior within the
class of systems with long-range interactions \cite{art,AntoniTorcini,rev1}.
For $A=0$ it reduces to the standard HMF model [Eq.~\eqref{hmf1}]. More
generally, while the standard HMF model shows a continuous transition from the
homogeneous to the ferromagnetic, single-cluster phase, the generalized HMF
model can exhibit both continuous and discontinuous canonical transitions
 depending on
the value of $A$, as shown in its phase diagram (\figref{FigPhaseDiagram2d}).
Moreover, there exists another ordered phase, called hereafter the
double-cluster phase, which is composed of two clusters of oscillators
separated on average by $\pi$ both in $x_i$ and $y_i$,
 and thus characterized by
finite values of $P$ and a vanishing magnetization $M$. On the
microscopic side, the essential difference with the standard HMF model is that
here a single oscillator has four variables, and hence can be chaotic
in the absence of any coupling with either an external field or other
oscillators. The largest LE of the full system is therefore not
purely determined by the coupling effect, unlike in the standard HMF model, but
receives also a contribution from the local dynamics, which depends on the
single-oscillator energy. 

\begin{figure}[t]
\begin{center}
\includegraphics[width=\hsize,clip]{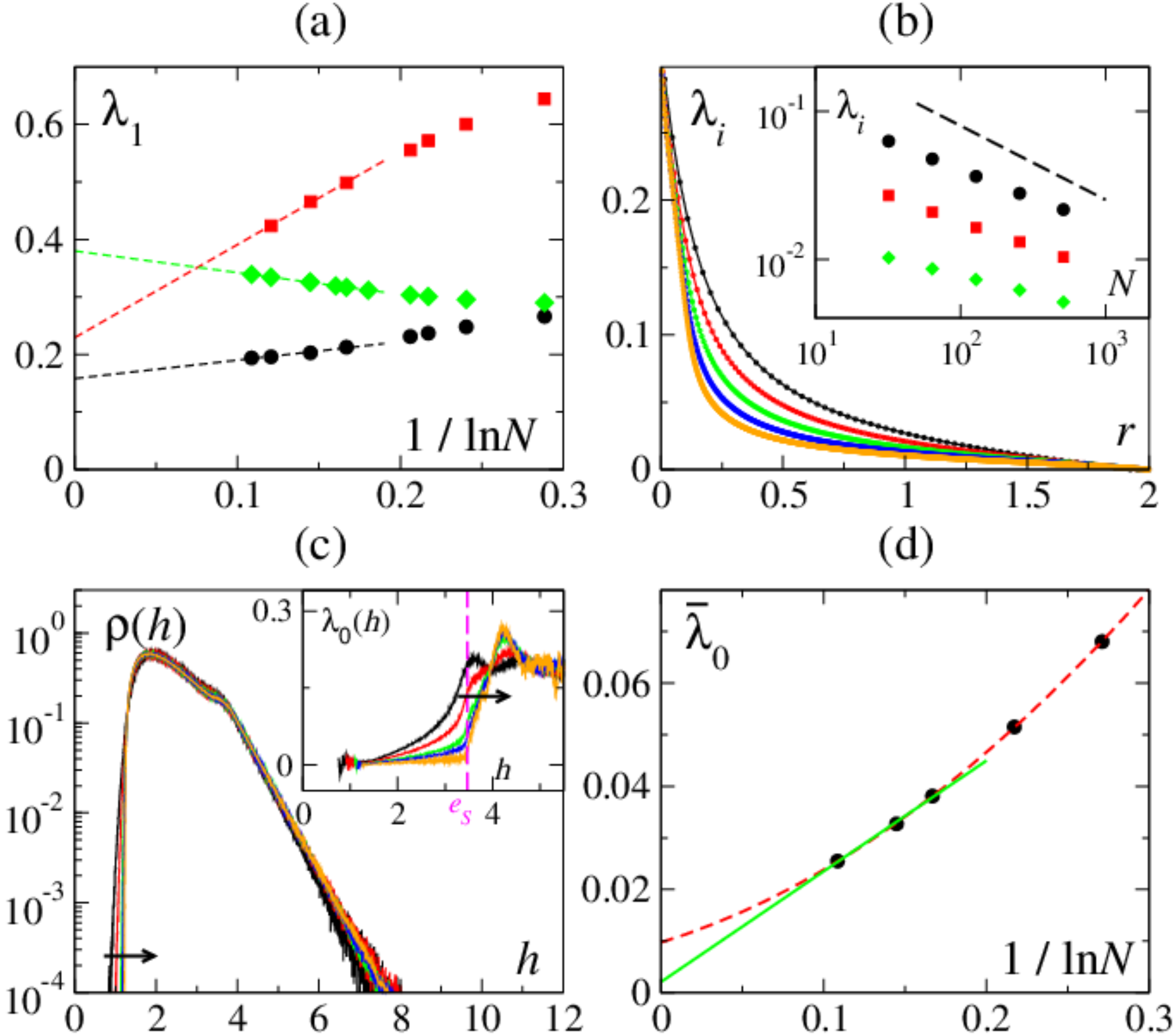}
\caption{(color online). Lyapunov exponents in the generalized HMF model. (a) Largest LE of the full system, 
$\lambda_1$ vs $1/\ln N$ for $A=0.2, U=1.4$ (black circles);
 $A=1.0, U=1.5$ (green diamonds); and $A=6.0, U=5.0$ (red squares).  The dashed lines indicate linear fits to the scaling regime. (b)
Full spectrum $\lambda_i$ vs $r = (i-0.5) / N$ with $N=32, 64, 128, 256$, and $512$ (from top to bottom) for $A=1.0, U=1.5$. Inset:
$\lambda_i$ vs $N$ for fixed rescaled indices $r = 0.5, 1.0, 1.5$ from top
to bottom. (c) Distribution of the single-oscillator energy $h$ for
$A=1.0, U=1.5$ and systems of size $N = 40, 100, 400, 1000$, $10000$. The arrow
indicates increasing system size. In the inset: energy-dependent
single oscillator LE for the same oscillator numbers $N$. The vertical
dashed line marks the position of the mean field saddle energy $e_S = 2+ A (P+1)$, with
the numerical estimate $P=0.469(1)$. 
 (d) The LE $\bar\lambda_{0}$ of a single oscillator forced by the
 full system of size $N$ for $A=1.0, U=1.5$. The red dashed and green solid lines show
 the results of a quadratic and a linear fitting, respectively, which result in finite asymptotic values of $\bar\lambda_{0}$.}
\label{fig-HMF2d}
\end{center}
\end{figure}

Figure \ref{fig-HMF2d}(a) shows the largest LE $\lambda_1$
measured for three different sets of parameter values: 
$A=0.2, U=1.4$ (point $P_1$ in Fig.~\ref{FigPhaseDiagram2d}), in the
single-cluster phase, close to a (canonical) continuous transition
[black circles in Fig.~\ref{fig-HMF2d}(a)]; $A=1.0, U=1.5$ (point $P_2$),
 in the single-cluster phase, close to
a (canonical) discontinuous transition (green diamonds);
and $A=6.0, U=5.0$ (point $P_3$), in the double-cluster phase (red squares).
In all the three cases, the largest LE
 shows the $1/\ln N$ scaling toward nonzero
asymptotic values, similarly to the standard HMF model. 
From the figure, one notices that in proximity of continuous
transitions $\lambda_1$ decreases with $N$, while at $P_2$,
in proximity of the discontinuous transition, the maximal LE
increases with the system size. This peculiarity deserves 
further investigations.

Analogously to the 1D case, it is instructive to start comparing with the
behavior of a single oscillator forced by constant order parameters.
At variance with the 1D case, the resulting value of the LE
can be positive here and depends on the energy. 
When one compares the extrapolated asymptotic values $\lambda_\infty$ 
 of the full-system LE with the maximum value
 of the single-oscillator LE, $\lambda_{M}$, over possible energy values,
 we obtain $\lambda_\infty = 0.16$ and $\lambda_{M}=0.11$ at $P_1$, 
 $\lambda_\infty=0.38$ and $\lambda_{M}=0.27$ at $P_2$,
 and $\lambda_\infty=0.23$ and $\lambda_{M}=0$ at $P_3$;
 the first asymptotic LE of the full system
 is systematically larger than $\lambda_{M}$. 
This is again a manifestation of the coupling pressure discussed in
Sec.~\ref{SecToy}.
For the first two cases, because of the chaotic dynamics of the single
oscillator, one does not need to introduce the two-family approach
 taken for the standard HMF model, but it
is more meaningful to refer to the treatment developed in Ref.~\cite{short}
 for dissipative systems, which predicts,
\begin{equation}
\label{eq:standard}
\lambda_\infty = \bar\lambda_{0} + \frac{D}{2}
\end{equation}
where $D$ is the diffusion coefficient
 for the fluctuations of the single-oscillator finite-time LE
 and $\bar\lambda_{0}$ is the single-oscillator LE
 with an appropriate averaging over energy values.
By taking into account this correction
 and using $\lambda_{M}$ instead of $\bar\lambda_{0}$ for the sake of simplicity,
we find: $\lambda_\infty \approx 0.18$  ($D=0.15$) at $P_1$, and
$\lambda_\infty \approx 0.33$  ($D=0.12$) at $P_2$. The new values are much
closer to the extrapolated ones, though there is still a remaining gap
(especially in the second case), which is presumably due to the fact that
Eq.~(\ref{eq:standard}) was derived under the assumption of short ranged
time correlations. The slow diffusion across different energy surfaces makes
this assumption at least questionable.
In contrast to these cases for the single-cluster phase,
 the situation in the double-cluster phase is rather analogous
 to the standard HMF model; because $M=0$ in the infinite-size limit,
 the equations of motion in this limit reduce to
\begin{equation}
 \dot{p}_{x,i} \pm \dot{p}_{y,i} \simeq -A P\sin(x_i \pm y_i - \psi_\pm),
 \label{HMF2dDoubleCluster}
\end{equation}
 which are equivalent to two uncoupled standard HMF models
 [Eq.~\eqref{model1}].
However, our theoretical approch in Sec.~\ref{SecToy} should not be applied
 directly to this case, because it deals with finite sizes, where
 the two variables in Eq.~(\ref{HMF2dDoubleCluster}) are coupled
 in a non-trivial manner.
%

We also studied the full Lyapunov spectrum $\lambda_i$
 of the generalized HMF model.
The results shown in
\figref{fig-HMF2d}(b) are obtained at $P_2$. They indicate that the full
spectrum becomes flatter and flatter for larger sizes, with an apparent
power-law decay of $\lambda_i$ with fixed rescaled index $r$ (inset).
While the convergence toward zero was expected in the 1D case, 
this behavior is questionable for the 2D model since 
in this case the single oscillator may be chaotic in the presence of a constant
magnetization.
In fact, it is reasonable to expect that, analogously to the dissipative
mean-field models discussed in Ref~\cite{short},
 the exponents in the bulk of the spectrum converge
 to the value of the LE of a single forced oscillator
 without coupling in tangent space.
However in a Hamiltonian model such as the 2D HMF,
it is not clear which exponent one should refer to, as it depends on the
energy and, moreover, the phase space is filled with stable islands.
Direct measurement of the energy of a single oscillator
 in the full system of size $N$ indicates that,
 within sufficiently long time scales,
 the single-oscillator energy diffuses as largely
 irrespective of $N$ [\figref{fig-HMF2d}(c)].
Given this existence of a well-defined distribution function $\rho(h)$
 for the single-oscillator energy, an appropriate reference value
 $\bar\lambda_{0}$ for the single-oscillator LE would be
 simply the exponent
 averaged with this distribution function, namely,
\begin{equation}
\bar\lambda_{0} = \int dh \rho(h) \lambda_{0}(h)
\end{equation}
where $\lambda_{0}(h)$ is the energy-dependent single-oscillator
LE. The data reported in the inset of Fig. \ref{fig-HMF2d}(c) indicate
that in the thermodynamic limit $\lambda_{0}(h)$ 
vanishes for $h < e_S= 2 + A (P +1)$ ($e_S$ being the single oscillator
saddle energy in the mean field limit), while finite contributions arise for larger energy values.
At finite $N$, $\bar\lambda_{0}$ is nothing but the conventional time-averaged LE
 of a single forced oscillator and is reported in \figref{fig-HMF2d}(d).
This substantially decreases with increasing $N$
 and, in particular, in the infinite-size limit,
 it can reach a positive but quite small value
 of the order of $10^{-3}$
 [estimated by a linear fit in \figref{fig-HMF2d}(d)].
This indicates that the decreasing bulk exponents
 reported in the inset of \figref{fig-HMF2d}(b)
 can have such a small but positive asymptotic value,
 which is however indistinguishable from zero
 from the available numerical data.
Moreover, one should notice that each oscillator has two nonnegative LEs;
 the first one can be positive or zero as already discussed,
 while the second one is always zero because of the continuous time.
It implies that one may even expect the occurrence of two bands
 in the asymptotic bulk spectrum, in correspondence
 with these two single-oscillator LEs.
Further studies are necessary to clarify these issues,
 to elucidate the generality of the extensivity and subextensivity
 found for the standard HMF model.

\section{Conclusions and open problems}
\label{sec:conclusions}

The question whether the largest LE in the HMF model remains positive
 or converges to zero in the thermodynamic limit has remained unsettled
for a long time. We have shown here that the largest LE is indeed positive
 by making use of 
several subtle properties of globally-coupled systems. The
first ingredient is what we call the ``coupling pressure" which induces
a finite increase in the largest LE (with respect to the
LE of a single oscillator forced by the mean field).
Coupling pressure is a general phenomenon that occurs in globally-coupled models
of both dissipative and conservative dynamical systems and arises from the
fluctuations of single-oscillator finite-time LEs. 
However, the 1D HMF dynamics is
even more subtle, since the LE of the single oscillator under a constant field
 is strictly zero, and its relevant fluctuations must be computed by
referring to a special type of trajectories that come close to the separatrix.
More ``natural'' is the behavior of the 2D HMF, since the single oscillator
dynamics is chaotic and thus the overall scenario is analogous to that
of standard dissipative chaotic systems (see Ref.~\cite{short}).
Altogether, our results indicate that the thermodynamic limit is rather 
singular. If one first takes the limit $N\to\infty$, no fluctuations
can be expected and no signature of chaos detected. On the other hand, we have
shown that the largest LE of an arbitrarily large system
is always positive. This means that representations of the dynamics such as
that built in the Vlasov equation (which corresponds to assuming $N=\infty$)
lose the chaoticity of the original dynamics captured by the largest LE.

On a more quantitative level, we have been able to derive an explicit expression
for a lower bound of the largest LE. It would be interesting to improve
the argument to determine a more accurate estimate and possibly predict the
dependence on the energy (or, equivalently, the temperature). Our numerical
analysis indicates that the largest LE stays indeed
 positive in the ordered phase.
Possibly, it can remain positive in the limit $U \to 0$,
 but a purely numerical approach is out of question
 because one would need to simulate large enough systems to guarantee
 the presence of some oscillators near the saddle
of the corresponding potential. 
Near $U=0$, the probability for an oscillator
to come close to the separatrix goes to zero, and thus
simulations are utterly unfeasible.

Although we have not been able to extend the theoretical arguments beyond the
largest exponent, we have undertaken also a general investigation of the entire
Lyapunov spectrum to investigate the extensivity of the chaotic dynamics.
Our numerical analysis suggests  that the asymptotic number
 of unstable directions is not extensive
(it grows probably like $\ln N$). It would be desirable to develop some even approximate
argument to justify this scaling behavior, which is, so far, only based on
numerical simulations.

Finally, we have also analyzed the Lyapunov spectrum for the 2D HMF. Such a system
is less pathologic than the 1D model, since the single oscillator dynamics
is chaotic and it is therefore obvious to expect positive LEs.
However, a problem remains to be settled regarding
the scaling behavior of the full spectrum. On the basis of all arguments
developed here and in Ref.~\cite{short}, we would expect that the 
bulk of the Lyapunov spectrum (at least for $r<1$) converges to a finite value. 
However this is not yet seen in our simulations. We cannot exclude that this 
is because the finite value associated to the single oscillator dynamics is really small.

All in all, the results presented here need to be put of firmer ground by more 
rigorous mathematical approaches, especially since we have shown that strong finite-size 
effects are at play. The open questions mentioned above also require further work. 
It is our hope that the rather subtle phenomena uncovered here will attract such needed attention 
in the future.

\acknowledgments
We thank M. Antoni for providing us the figure of the
phase diagram of the generalized HMF model and S. Ruffo
for useful discussions. AT thanks X. Leoncini for fruitful
exchanges of information.



\begin{thebibliography}{99}
\bibitem{noteJ} Here we adopt the standard normalization for the
  extensive interaction introduced by Antoni and Ruffo, which is
  different from the ones used in \cite{J-hmf}.
\end{thebibliography}
\end{document}